\documentclass[11pt,letterpaper]{article}

\usepackage[T1]{fontenc}
\usepackage[english]{babel}
\usepackage{times}

\usepackage[letterpaper,margin=1in]{geometry}

\usepackage{amsmath,amstext,amsbsy,amsopn,amsthm,amsfonts,amssymb}
\usepackage{mathtools}
\usepackage{url}
\usepackage{tabularx,booktabs}
\newcolumntype{C}{>{\centering\arraybackslash}X}
\usepackage{booktabs,array,colortbl}
\usepackage[]{graphicx}
\usepackage{rotate}
\usepackage{wrapfig}
\usepackage{caption}
\usepackage[caption=false]{subfig}
\usepackage{lastpage}

\usepackage[small]{titlesec}
\usepackage{paralist}

\usepackage{verbatim} 
\usepackage[usenames,dvipsnames,svgnames]{xcolor}
\usepackage{multirow}
\usepackage{algorithmic}
\usepackage[linesnumbered,ruled,vlined]{algorithm2e}

\usepackage{fancyhdr}
\providecommand{\keywords}[1]{\textbf{\textit{Keywords---}} #1}

\renewenvironment{thebibliography}[1]{%
	\begin{oldthebibliography}{#1}%
		\setlength{\parskip}{0.0cm}%
		\setlength{\itemsep}{0.0cm}%
	}%
	{%
	\end{oldthebibliography}%
}

\begin{document}

\title{\textbf{PoAh}: A Novel Consensus Algorithm for Fast Scalable Private Blockchain for Large-scale IoT Frameworks}

\maketitle


\author{
	\begin{center}
		\begin{tabular}{cc}
		Deepak Puthal &	Saraju P. Mohanty  			\\
	School of Computing &		Dept. of Computer Science and Engineering 	  \\
	Newcastle University, UK &		University of North Texas, USA.			  \\
	Email: \texttt{Deepak.Puthal@newcastle.ac.uk} &		Email: \texttt{saraju.mohanty@unt.edu}    \\\\
		\end{tabular}
		\begin{tabular}{cc}
	Venkata P. Yanambaka  &		Elias Kougianos  \\
	School of Science and Engineering &		Department of Electrical Engineering  \\
	Central Michigan University, USA. &		University of North Texas, TX 76203.    \\
	Email: \texttt{yanam1v@cmich.edu} &		Email: \texttt{elias.kougianos@unt.edu}   \\
			\\\\
		\end{tabular}
	\end{center}
}

\cfoot{Page -- \thepage-of-\pageref{LastPage}}

\begin{abstract}
In today's connected world, resource constrained devices are deployed for sensing and decision making applications, ranging from smart cities to environmental monitoring. Those recourse constrained devices are connected to create real-time distributed networks popularly known as the Internet of Things (IoT), fog computing and edge computing. The blockchain is gaining a lot of interest in these domains to secure the system by ignoring centralized dependencies, where proof-of-work (PoW) plays a vital role to make the whole security solution decentralized. Due to the resource limitations of the devices, PoW is not suitable for blockchain-based security solutions. This paper presents a novel consensus algorithm called Proof-of-Authentication (PoAh), which introduces a cryptographic authentication mechanism to replace PoW for resource constrained devices, and to make the blockchain application-specific. \textcolor{black}{PoAh is thus suitable for private as well as permissioned blockchains}. Further, PoAh not only secures the systems, but also maintains system sustainability and scalability. The proposed consensus algorithm is evaluated theoretically in simulation scenarios, and in real-time hardware testbeds to validate its performance. Finally, PoAh and its integration with the blockchain in the IoT and edge computing scenarios is discussed. The proposed PoAh, while running in limited computer resources (e.g. single-board computing devices like the Raspberry Pi) \textcolor{black}{has a latency in the order of 3 secs}.
\end{abstract}



\section{Introduction}
\label{SEC:Introduction}

The Internet of Things (IoT) has various definitions across the research community. A network connecting various devices called ``\textit{Things}'' that are responsible for various tasks not limited to collecting data from the environment, is called ``Internet of Things''. The devices referred as ``\textit{Things}'' can be any object, sensor, human or other device. According to various definitions of the IoT, for a device in the network to be considered a ``\textit{Thing}'', it should have communication capabilities. With their advantages, different IoT architectures have become a backbone for a wide variety of applications ranging from smart healthcare to industrial IoT and smart cities \cite{Castanho_TEM_2019, Corbett_TEM_2017-May}. Specifically in a smart city, millions of sensors and things are critical for the IoT which build its various components \cite{Hsiao_TEM_2019, Mohanty_CEM_2016-Jul}. All these frameworks will have security requirements which this work envisions to be addressed through the use of blockchain technology.

\textcolor{black}{Based on the IoT architecture, individual devices are identified with their own ID in a network and it is not necessary to follow the TCP/IP communication protocol stack. There are several other communication protocols for lightweight and low bandwidth systems to build to-end communication in an IoT network \cite{dorri2017blockchain}. One of the major concerns in an IoT architecture is data collection, while processing is performed at a lower level which uses low performance devices such as single board computers. Collecting environmental variables or data can be done using middleware for better performance in collection and transmission \cite{Moreno_TII_2017-Apr}. In a server-based scenario, an edge layer is introduced for near real-time data processing \cite{shi2016edge}, which includes Edge Datacenters (EDCs). These EDCs are not centralized and hence do not have any dependencies, which makes the IoT architectures suitable for emergency data processing \cite{zanella2014internet,puthal2016threats}. The integration of IoT and distributed edge networks is a good fit for the application oriented deployment.}

With the help of Edge Datacenters, IoT architectures can be designed with such resource constrained and low performance devices which are capable of low power consumption. Such architectures aid various mission-critical applications including military, industrial IoT, event detection. In such applications, attaining real-time data processing and secure data transfer are of paramount importance. Of all the attributes of such environments, security should be given highest priority. Research is already actively being conducted and many algorithms were developed for this purpose which involves cryptography \cite{minoli2017iot, lee2017security}. In the two categories of cryptography, symmetric cryptography gives a 1000 times increase in performance compared to asymmetric cryptography and this is one of the reasons it is widely adopted for IoT architectures \cite{xu2014security,puthal2017dlsef}. \textcolor{black}{For both of these cryptography approaches, a central entity plays a vital role during the security initialization phase, such as acting as a key distribution center for symmetric cryptography and acting as the certificate authority for asymmetric cryptography. A failure at this single point can lead to a catastrophic failure in the entire system and compromise the entire network.} 

\begin{figure}[htbp]
	\centering
	\includegraphics[width=0.70\textwidth]{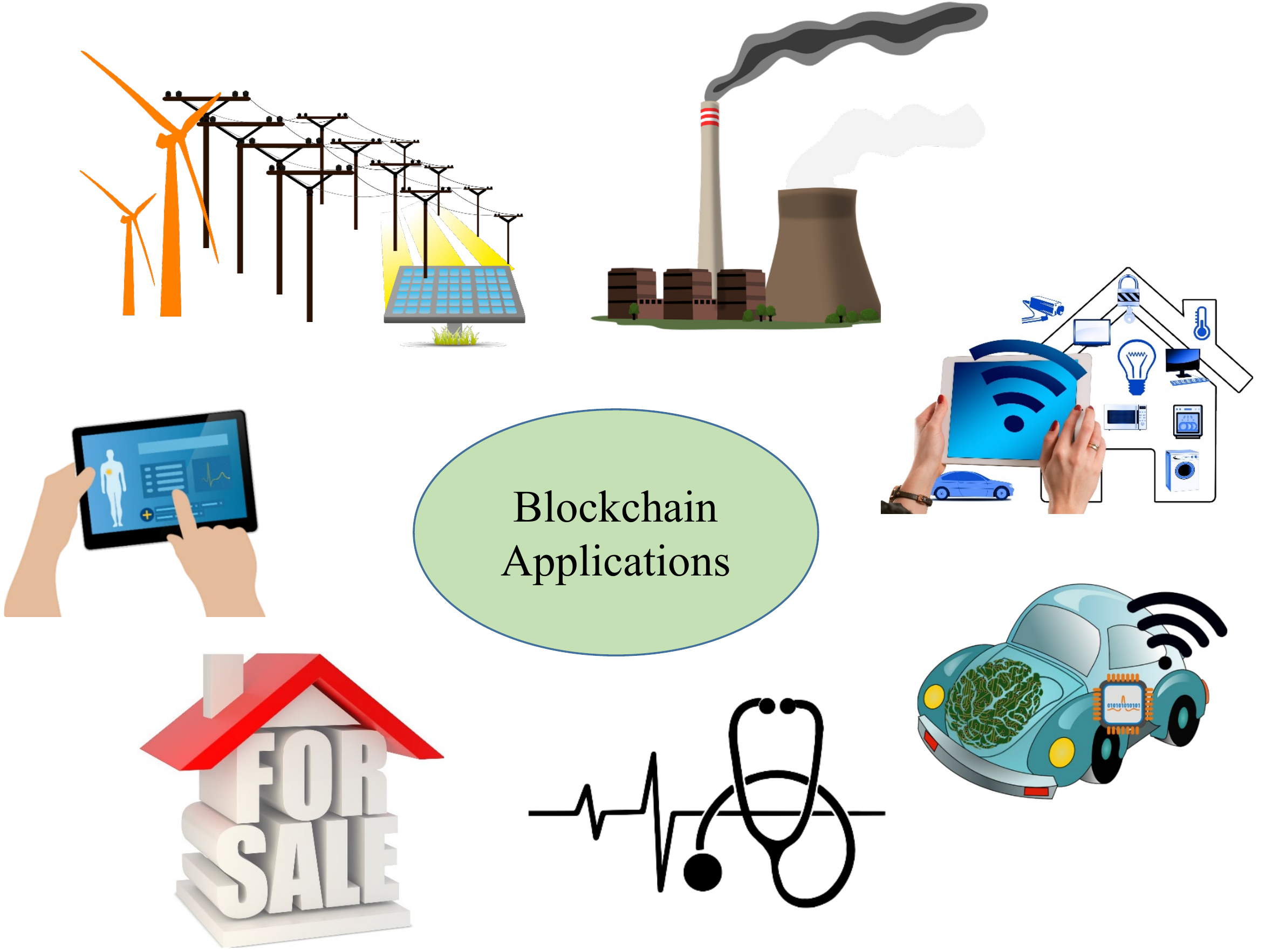}
	\caption{Applications of blockchain.}
	\label{fig:Blockchain_Applications}
\end{figure}

The requirement of a central entity to organize and perform cryptographic operations or other tasks can be achieved with the implementation of a blockchain. \textcolor{black}{In fact, the blockchain's primary contribution is the resolution of the age-old problem of achieving a consensus via Proof-of-Work authentication}. The blockchain uses a decentralized public ledger for organizing the data and executing transactions. \textcolor{black}{The ledger is decentralized in nature, such that individual peers in the network possess a copy of the entire or part of the ledger, depending on its architecture, after block validation by the miners.} A blockchain is cryptographically anchored and tamper-proof and is a record of the different transactions that occurred among the participants in the network \cite{puthal2018blockchain}. 
There are various applications of the blockchain as shown in Fig. \ref{fig:Blockchain_Applications}.

\textcolor{black}{Blockchain technology replaces a centralized entity with consensus algorithms among the participants to secure the system in a decentralized fashion. This consensus allows the participants to trust each other and complete the transfer or transactions among the participants \cite{zyskind2015decentralizing}.} All the blocks are connected to the previous blocks with the help of cryptographic hash algorithms and hence the transactional values cannot be tampered with once the block has been added to the blockchain. Proof-of-Work (PoW), Proof-of-Stake (PoS) and others are some of the consensus algorithms that are implemented in the blockchain. However, one of the major concerns of PoW and PoS is the cost of calculation. They require high computational power and energy which makes them unsuitable for IoT applications \textcolor{black}{that choose the blockchain as a means of decentralization}. Hence this paper proposes a new algorithm, Proof-of-Authentication (PoAh),  which can be incorporated into resource constrained distributed systems. \textcolor{black}{To the best of the authors' knowledge, this is the first attempt towards a lightweight blockchain consensus approach for scalable IoT deployment.}

The rest of the paper is organized in the following manner: Section \ref{Sec:Contributions} presents the contributions of this work. A brief overview of blockchain technology and relevant research is presented in Section \ref{Sec:StateOfTheArt}. The proposed novel Proof-of-Authentication (PoAh) algorithm is discussed in Section \ref{Sec:Proposed_PoAh}. A brief discussion on the analysis of the proposed consensus algorithm is presented in Section \ref{Sec:PoAh_Analysis}.
Experimental results are presented in Section \ref{Sec:Experiments}. Conclusions and future directions are summarized in Section \ref{Sec:Conclusions}.

\section{Contributions of this Work}
\label{Sec:Contributions}

\subsection{Problem Definition}

\begin{figure*}[htbp]
	\centering{\includegraphics[width=0.95\textwidth]{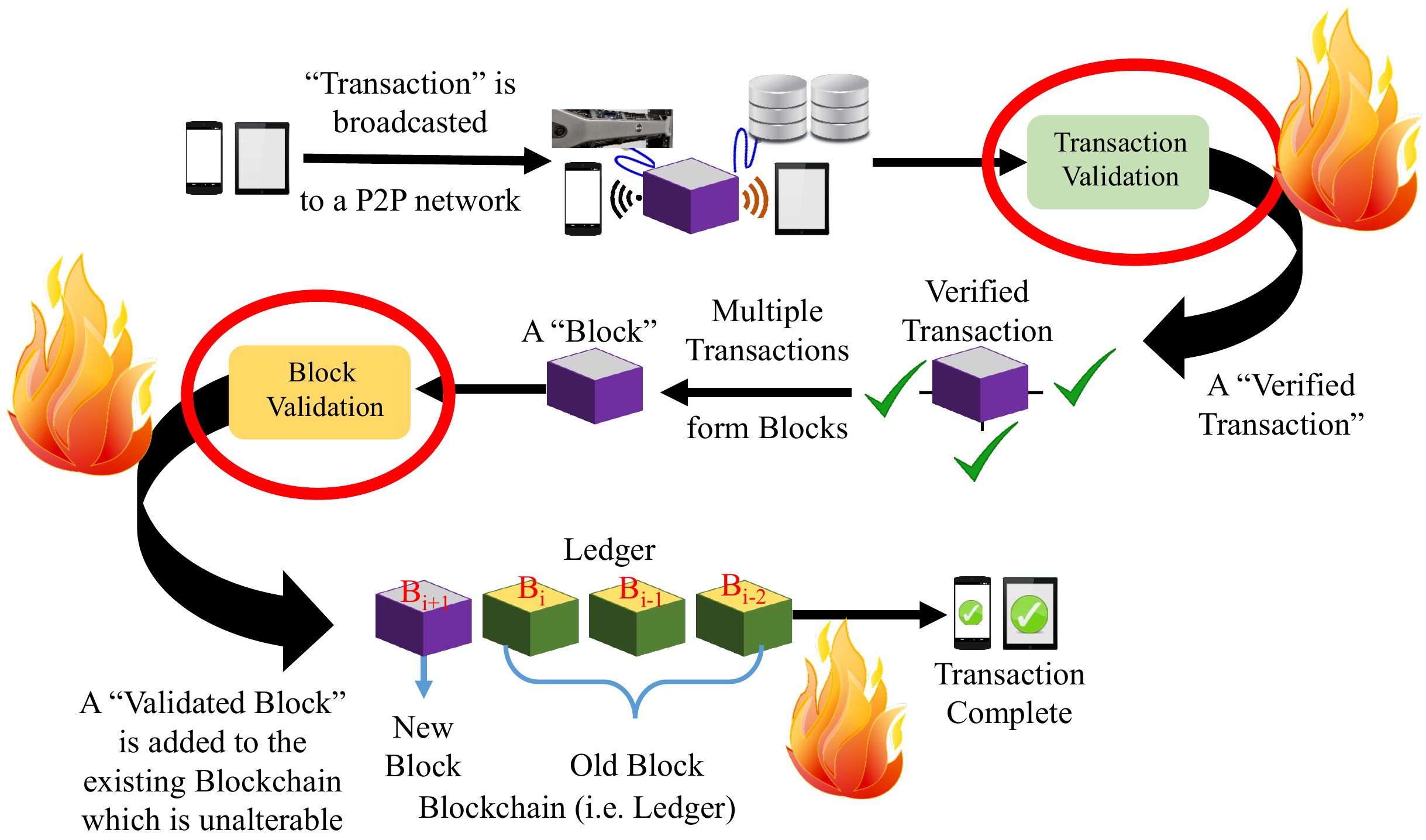}}
	\caption{Process of block creation and addition into the ledger. Flames are used to graphically highlight computationally expensive processes.}
	\label{FIG:Blockchain}
\end{figure*}

Fig. \ref{FIG:Blockchain} shows the process of generation, validation and addition of a block to the blockchain and the issues present in the blockchain. Once a transaction is initiated, it is broadcast to a network of devices. \textcolor{black}{Each device in the network will have a copy or part of the ledger in its respective local storage. Whether to maintain a copy of the entire ledger, or only the recent blocks is determined by the node administrators based on storage and power considerations. However, miner nodes always contain the entire blockchain.} The miners will then validate each transaction and create a block comprising of multiple transactions, as shown in Fig. \ref{FIG:Blockchain}. The transaction validation and block validation consume energy as they require the processing power from various devices. Then, the validated block gets added to the blockchain. The local storage of the devices also becomes a bottleneck as over time, a big block size consumes a lot of memory which is limited in IoT devices. 

It is estimated that by 2022 there will be 18 billion devices spread across the global network, with the majority of them being smart devices and small sensors \cite{novo2018blockchain}. The widespread adoption of the IoT brings forward new technological challenges in data privacy and security \cite{huang2017decentralized}.

The most promising solution to this issue is the implementation of blockchain technology into IoT networks. Blockchain is the technology that has maintained the integrity and security of cryptocurrencies since the introduction of Bitcoin in 2009 \cite{nayak2017blockchain}. This technology is predicted to be able to:
\begin{itemize}
   \item \textcolor{black}{Maintain transaction data integrity, authentication and immutability as required in IoT networks \cite{nayak2017blockchain}. }
    \item Maintain the integrity of stored data within IoT networks, making the data difficult to modify \cite{kshetri2017can}.
    \item Increase reliability of IoT networks through decentralization of nodes \cite{nayak2017blockchain}. 
\end{itemize}

However, general IoT systems have different requirements than cryptocurrencies. Some everyday tasks and applications cannot be solved with a blockchain implementation \cite{Wang_ACM_2019}. Various issues must be resolved for successful integration of the blockchain into such systems. These can include:
\begin{itemize}
    \item Reducing the time to add blocks to the blockchain to improve system responsiveness \cite{xin2017scaling}.
    \item Improving physical Internet infrastructure to support increased Internet usage of decentralized networks \cite{kuzmin2017blockchain}.
    \item Accounting for power, CPU and memory constraints of less powerful computers.
\end{itemize}

Blockchain implementation into the IoT is a field that is currently receiving a lot of attention from the academic and industrial communities due to its promise, with many researchers designing new architectures and solutions to implement the blockchain into IoT \cite{ouaddah2016fairaccess,novo2018blockchain,Biswas_JIoT_2020}.

The blockchain promises to be a possible solution to the privacy and security issues posed by IoT systems. The primary goal of this research project is the creation of an original IoT architecture, with blockchain implementation to secure user access to data and the system.

This paper focuses upon the utilization of a private blockchain network in an IoT system. This is due to several advantages that private blockchains have over public blockchains, when implemented into IoT systems. Private blockchain networks generally have lower network delay. This is due to the measures public blockchains have to utilize to motivate peers. In addition there are also fewer issues of trust, resulting in lower security measures required between nodes \cite{xin2017scaling}. This leads to a more responsive network, which may be desirable for some implementations of the IoT. Another advantage of a private blockchain implementation is that system data are contained entirely within the network, which allows network owners full control of their personal data \cite{xin2017scaling}.

\textcolor{black}{There are always security threats in the IoT from both inside and outside attackers \cite{puthal2016threats, minoli2017iot, lee2017security}. Considering the available cryptographic security solutions, both symmetric and asymmetric cryptography require a centralized dependency for initialization. Therefore, one point failure can lead to compromise of the entire system. To overcome such possible hazards, the blockchain provides a decentralized framework to secure the system. The problem with existing consensus algorithms is that they are not scalable for IoT systems with a very large number of nodes. We address this problem in this work and we are proposing a novel scalable blockchain consensus for the IoT. The  proposed consensus approach is described in the following section.} 


\subsection{Proposed Novel Solution}

Integrating the blockchain with the IoT is still a challenging task due to the resource constrained nature of IoT devices. At the same time, the IoT system demands a decentralized security solution due to the distributed and untrusted environment. To address these issues, this paper implements a novel blockchain consensus algorithm called PoAh for IoT systems.  The novel aspects of the proposed consensus algorithm towards blockchain technology are listed as follows:
\begin{itemize}
    \item Proof-of-Authentication (PoAh) is introduced as a cryptographic authentication mechanism for lightweight blockchain. 
    \item PoAh is a new consensus algorithm and is validated for resource-constrained devices in IoT and edge computing.
    \item Finally, PoAh is validated theoretically and evaluated in both simulation and real-time hardware testbeds for performance.
\end{itemize}

\section{State of the Art of Blockchain Technology}
\label{Sec:StateOfTheArt}
\subsection{Scalability Issue of Proof of Work Based Decentralized Blockchain}

Bitcoin was the first implementation of the blockchain technology. It is a cryptocurrency which uses the concept of a decentralized distributed ledger at its core to remove the central entity for performing transactions between different participants in the network \cite{puthal2018blockchain}. Since then many other cryptocurrencies have appeared which use blockchain technology, such as Etherium which uses Smart Contracts \cite{shi2016edge}. The main intention of all these blockchain consensus algorithms and cryptocurrencies is the removal of the common entity which oversees the transactions. Every participant in the network will have a complete or partial copy of the ledger with the necessary transactions. They implement cryptographic hashes which allow them to be irreplaceable or irreversible once added to the blockchain. All the participants in the network use consensus algorithms for validating the transactions, and once a transaction is validated it is broadcast to every participant in the network and everyone updates the blockchain ledger \cite{zyskind2015decentralizing}. Along these lines, notwithstanding its remarkable strengths which incorporate decentralization, validation, transparency and immutability, the blockchain stretches security and privacy at all points in time.

\begin{figure}[htbp]
	\centering{\includegraphics[width=0.65\textwidth]{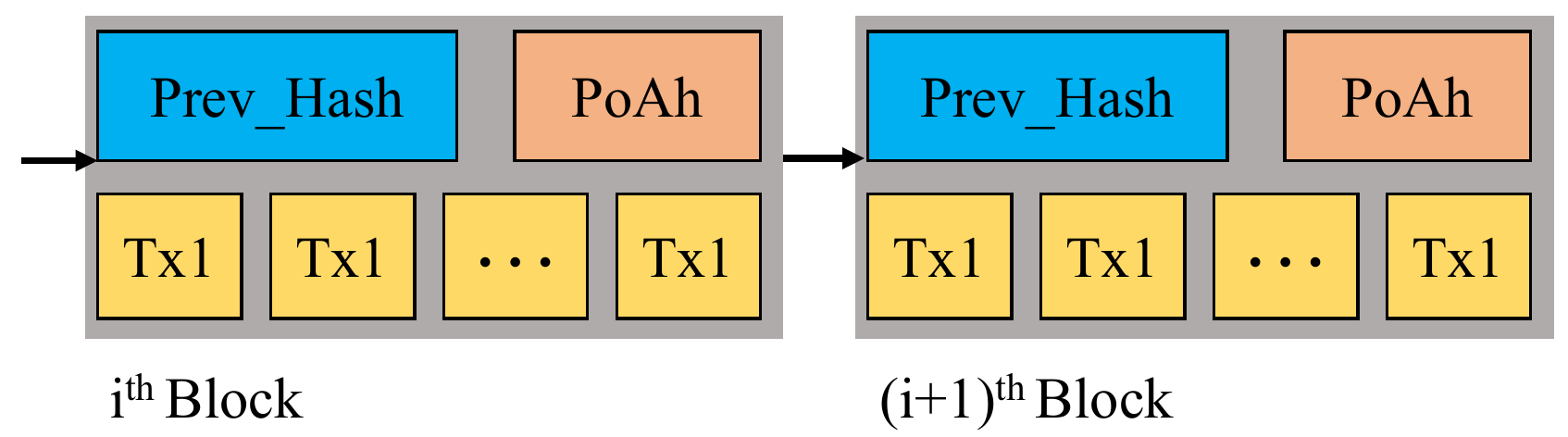}}
	\caption{Block processing with transactions and hashing.}
	\label{fig:trx}
\end{figure}

In distributed networks participants can be any kind of client, establishment or association sharing a copy of the ledger containing their real transactions in a progressive succession. A sequence of blocks  within in a ledger is shown in Fig. \ref{fig:trx}. Blocks are connected together with hash values in a sequential order for data immutability. Individual blocks comprise of sets of transactions, and individual transactions are digitally signed by the source and verified by other devices in the network before being added into the chain. Some features of the blockchain are now discussed.  

\textbf{Digital Signatures:} Individual nodes digitally sign the transactions and broadcast them to the network for verification by the miners of the network. Each block contains multiple transactions, and participants sign them if they wish to broadcast them to the network and include them to the blockchain. 

\textbf{Consensus:} The novelty of the blockchain is decentralized storage or management, where there are no central dependencies. The consensus algorithm plays a vital role in this operation to keep the transactions in a block and validate them for further processing. 

\textbf{Proof-of-work:} \textcolor{black}{This consensus algorithm searches for values generating a given cryptographic hash. This process essentially solves a cryptographic ``puzzle'' to validate blocks. As a result,  PoW provides greater security to the overall system. PoW requires intense resources i.e., electricity consumption is equivalent to 1.5 times the overall household electricity for one day in the USA \cite{zyskind2015decentralizing}.} \textcolor{black}{It is estimated that bitcoin transactions will consume close to the electricity in Denmark by 2020 \cite{king2012ppcoin} because of the computational power needed}.

\textbf{Cryptographic Hashes:} After successful proof-of-work, the block is broadcast to the network and all the participants compute a hash (such as SHA-256 for normal blockchain) to be used as the previous hash (``Hash\_Prev'' in Fig. \ref{fig:trx}) for the next block.

\subsection{Related Prior Research on Blockchain Scalability}

\begin{figure}[htbp]
	\centering{\includegraphics[width=0.85\textwidth]{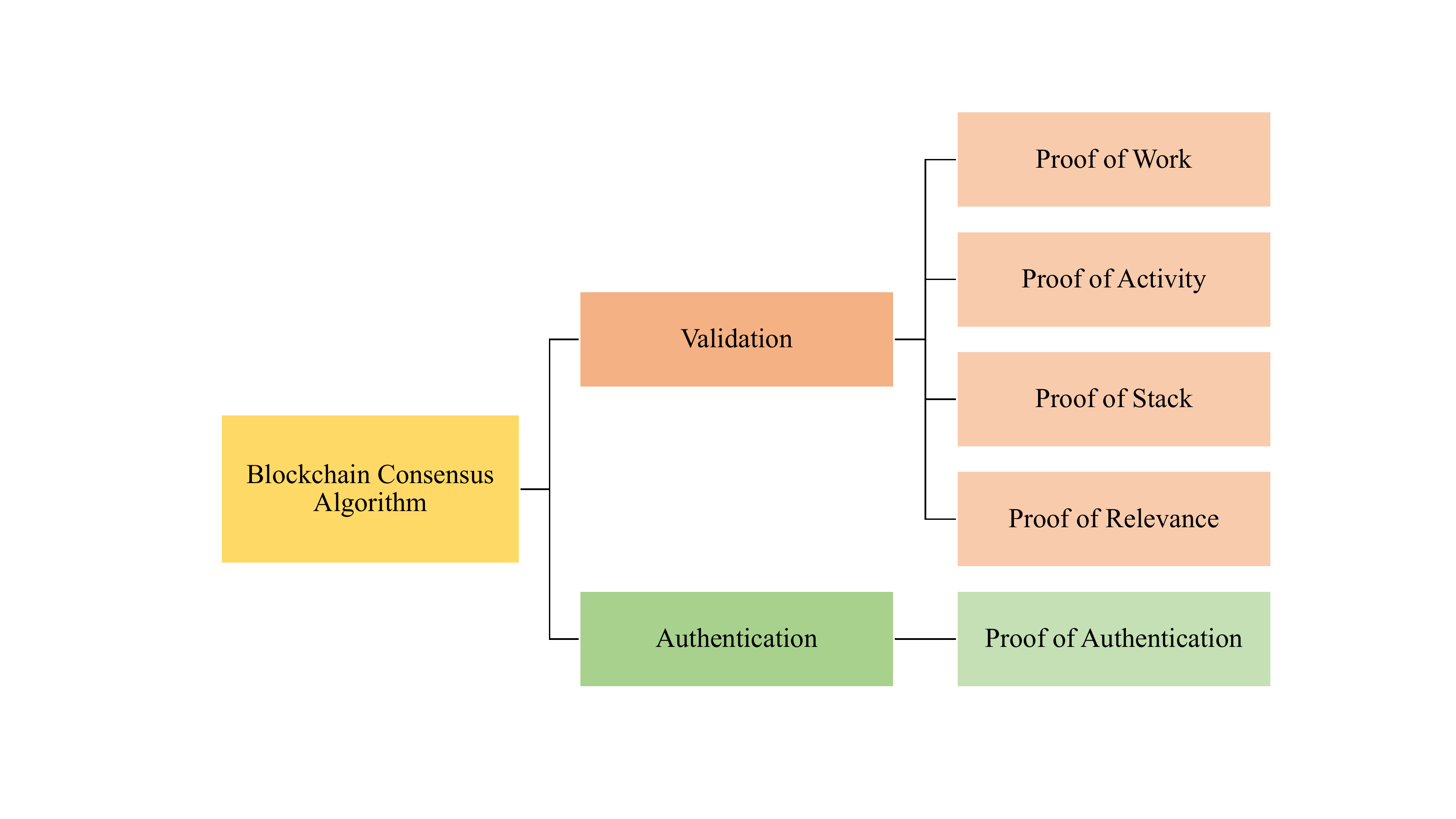}}
	\caption{Taxonomy of blockchain consensus algorithms.}
	\label{fig:tax}
\end{figure}

The blockchain can use several consensus algorithms, such as proof-of-stake, proof-of-work, proof-of-relevance and proof-of-activity. A lightweight consensus algorithm named Proof of Authentication has been briefly introduced in \cite{Puthal_Potentials_2019-Jan, Puthal_ICCE_2019_PoAh}. Fig. \ref{fig:tax} shows the various blockchain consensus algorithms and their classification.  The following section provides technical details about the proposed PoAh algorithm.

\subsubsection{Security Concerns in IoT Systems}

The major concern in the widespread adoption of IoT networks is the security of deployed systems \cite{huang2017decentralized}. The use of IoT networks is expected to grow in both the public and private domains and the abundance and value of data from both public and private IoT networks will attract attackers who will attempt the theft of this information. Within an insecure system, attackers can also potentially gain control of networked IoT devices which can have disastrous effects \cite{ouaddah2016fairaccess}. As an example, within a private home IoT network, an attacker can potentially gain control of lights, TV or connected cameras. Insufficient security can also lead to injury or even death when IoT systems become trusted with the control of larger machines, such as self-driving cars.

\subsubsection{Blockchain Implementation in IoT Systems}

Blockchain is the technology that enforces the integrity and value of Bitcoin and other cryptocurrencies. Since its introduction in 2008 by Satoshi Nakamoto (a pseudonym), the blockchain has proven to be reliable in protecting the value and integrity of cryptocurrencies \cite{nayak2017blockchain}. The secure properties of blockchain technology complement well with the security and privacy vulnerabilities of the current generation IoT systems \cite{kuzmin2017blockchain}.

Blockchain technology features the decentralization of data, which means control of the system is distributed amongst the peers within the network \cite{nayak2017blockchain}. This ensures that the user maintains control of their data as there is no central authority being sent gathered information from sensors. Decentralization of the data can also benefit the security of IoT networks as access control information is shared amongst all devices within the network \cite{ouaddah2016fairaccess}. If an attacker attempts to infiltrate a blockchain-secured IoT network, depending on the architecture of the system, more than half of the connected devices will need to be hacked simultaneously.

IoT systems with blockchain implementation is a new area of study, currently receiving significant research attention. The use of the blockchain in IoT systems has yet to be standardized. A detailed discussion on the advantages offered by the Blockchain technology when integrated into an IoT architecture and Smart Cities is presented in \cite{Sun_FI_2016}. The current challenge for engineers and researchers attempting to implement this system is the creation of functional architectures that can effectively merge these two technologies together, keeping the function of IoT systems while maintaining the privacy and security features of blockchain technology. A cross-chain framework which integrates multiple architectures of blockchain for robust and secure data management for IoT environments is presented in \cite{Jiang_Sensors_2019}. Similar implementations were also proposed in \cite{Sagirlar_archive_2018}, and \cite{Alphand_WCNC_2018}. A new blockchain based IoT architecture is proposed in \cite{Dorri_PerCom_2017} which can be deployed in a smart home, but the overhead added to the communication is high, which makes it unsuitable for various environments. A blockchain consensus algorithm which can authenticate the identity of the IoT device and maintain data protection was proposed in \cite{Dongxing_ICCCN_2018}.

\subsection{Existing Architectures of IoT Systems with Blockchain Implementation}

The fusion of blockchain and IoT technologies is still a new research topic and standards for implementing these two technologies have not yet been determined. There have been many architectures and proposed solutions for the implementation of the blockchain into the IoT and the design of these systems can vary significantly.

\subsubsection{Decentralized Access Control System}
Novo, in \cite{novo2018blockchain} details an IoT system which avoids the integration of IoT devices into the blockchain network and only the access control system of the network is managed and distributed by blockchain technology. This control system allows for lower power devices that cannot participate in the blockchain to be used within the system. 
To describe this design very simply, sensor networks collect data and send this information encrypted to the management hub. The management hub then transforms the information sent from the sensors into a format understandable by the rest of the blockchain network. The smart contract is secured with blockchain technology and is managed by public or private miners. It contains access control information which determines the devices that can access the sensor network controlled by the management hub.

\subsubsection{Fair Access Architecture}

The fair access framework \cite{ouaddah2016fairaccess} is an IoT system which allows a number of organizations or people to access each other's data. The access control of the data is managed by the same blockchain network, in which all members are connected to. Each member has a ``Wallet'' containing their access information and also contains all keys pertaining to the information they are allowed to view. To explain this system simply, if member A wants to request a resource that member B possesses, they would send their request and all related keys to the blockchain network. The blockchain network would then determine if member A has permission to access the resource form member B through mining. If member A is authorized to receive the information, then member B would transmit the data.

Blockchain implementation in IoT systems is a newly researched technology. While at the moment this technology seems promising, proper implementation can prove to be not viable or not even possible. This increases the risk of implementing prototype projects as the lack of support and preexisting modules available will significantly increase the difficulty of the task. Another problem with blockchain technology is that it is currently designed around mining cryptocurrencies and can take up to 12 minutes to validate a transaction, depending on the protocol used \cite{nayak2017blockchain}. While most delays are shorter, this waiting period can make the blockchain nonviable for some IoT systems that require fast response.

%

\section{The Proposed Novel Consensus Algorithm - Proof of Authentication}
\label{Sec:Proposed_PoAh}


The proposed Proof-of-Authentication is a new consensus algorithm proposed in this paper to build a lightweight and sustainable blockchain for resource-constrained devices. This consensus algorithm introduces an authentication mechanism during block validation. In other respects, it follows traditional communications.

\begin{figure*}[htbp]
	\centering
	\subfloat[\textcolor{black}{Proof-of-Work (PoW) consensus algorithm}]
	{\label{fig:PoW}\includegraphics[width=0.85\textwidth]{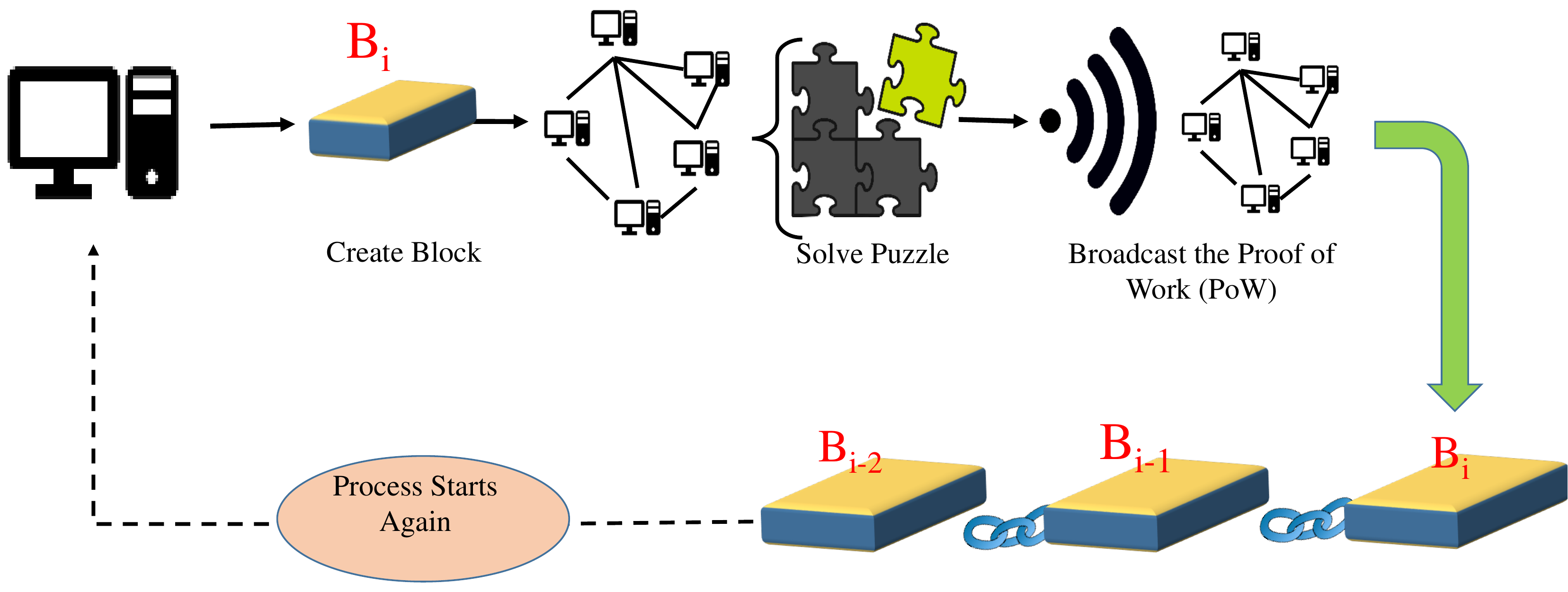}}\\
	\subfloat[\textcolor{black}{Proof-of-Stake (PoS) consensus algorithm}]
	{\label{fig:PoS}\includegraphics[width=0.85\textwidth]{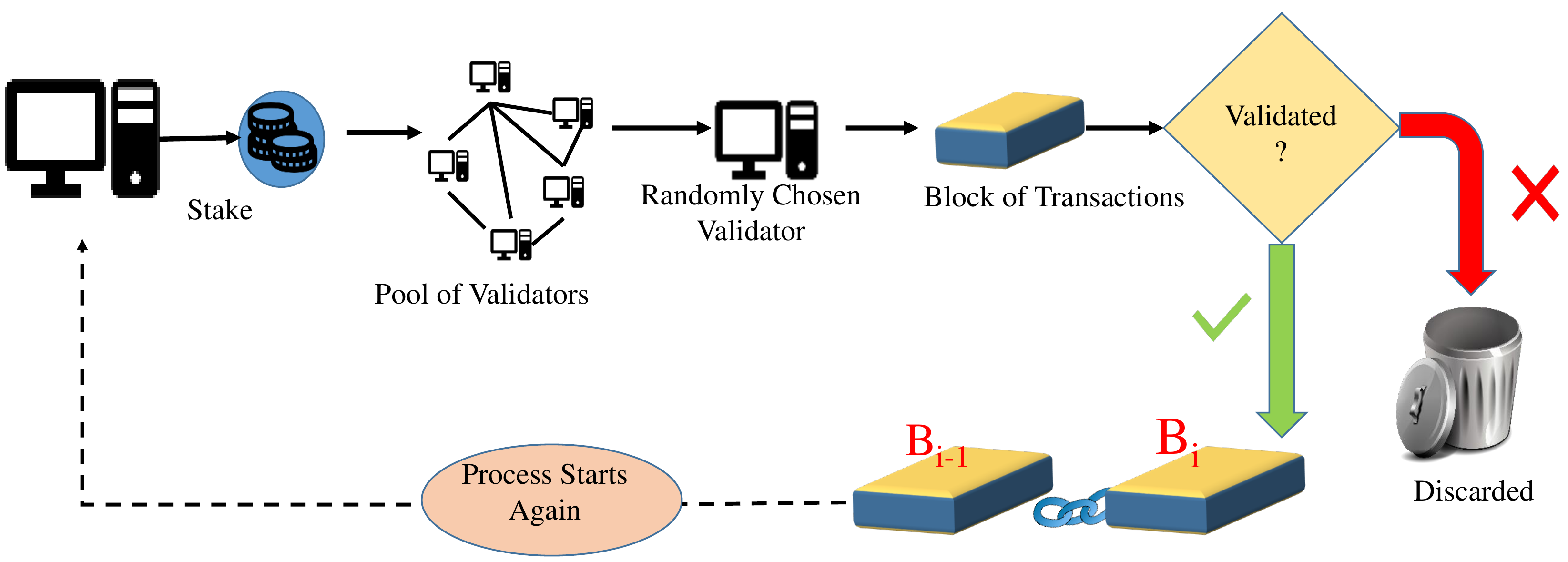}}\\
	\subfloat[The proposed Proof-of-Authentication (PoAh) consensus algorithm.]{\label{fig:PoAh}\includegraphics[width=0.85\textwidth]{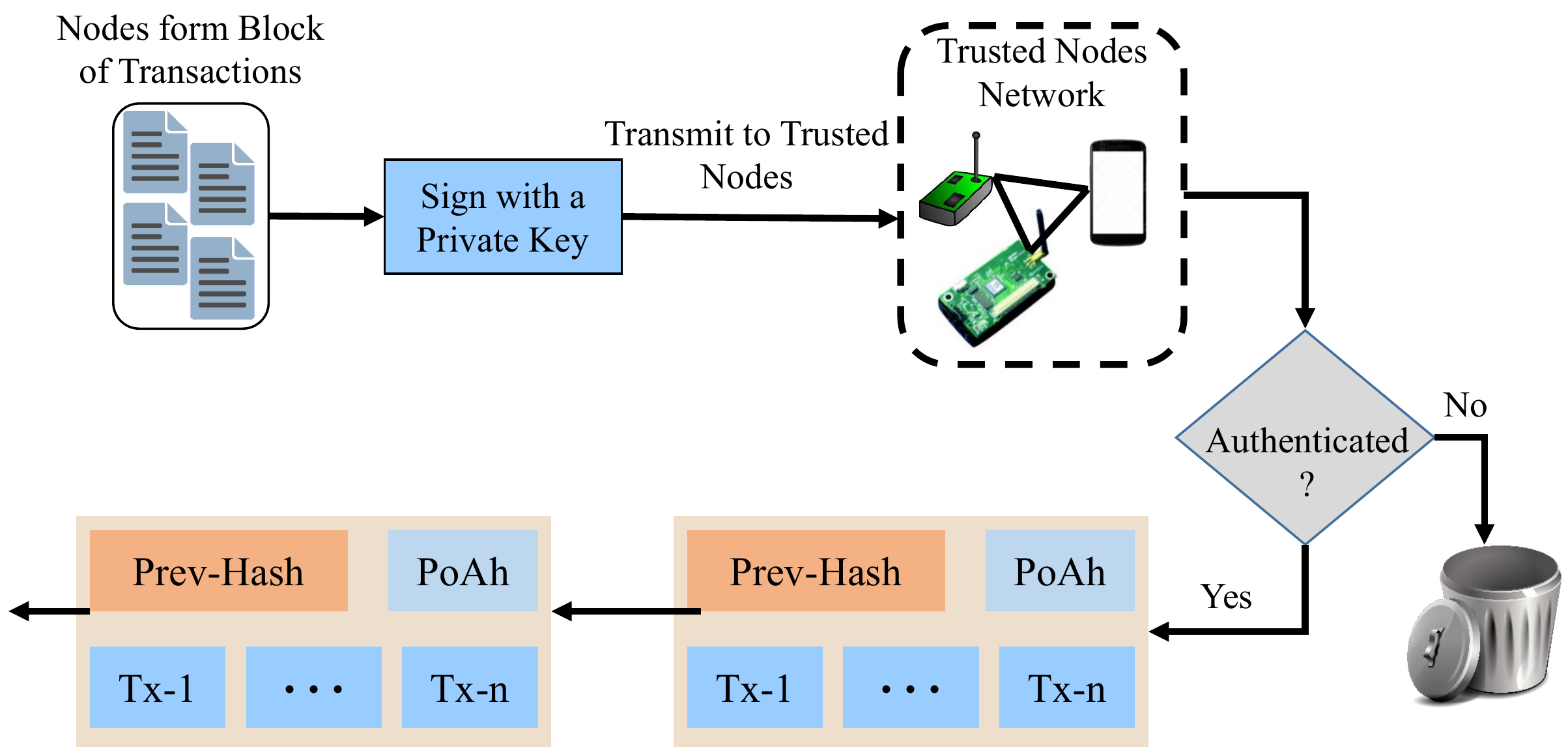}}
\caption{\textcolor{black}{Various consensus algorithms which can be used in a blockchain technology.}}
	\label{FIG:Various Consensus Algorithms}
\end{figure*}

Fig. \ref{FIG:Various Consensus Algorithms} provides a comparative overview of the proposed PoAh with the more common Proof-of-Work and Proof-of-Stake consensus algorithms. At the very beginning of the process, network precipitants generate transactions (Trx) with the sensed or collected data assembled to form a block. In the figure blocks are represented as $B = \{Trx1, Trx2, …, Trxn\}$. \textcolor{black}{The nodes broadcast the blocks for further evaluation and/or validation by trusted nodes in the network. We follow the standard IoT deployment steps to create the initial trusted node set based on geographical location. Trusted nodes are reachable from any part of the network.} The proposed model adopts the ElGamal crypto-system for encryption and decryption, i.e. $y=g^x(\mod p )$ where $x$ is the private key PrK and $y$ is the public key PuK. The large prime numbers for modulus operation $p$ and generator function $g$ are publicly known to all the network devices. Prior to block broadcast, the network user makes the public key PuK, i.e. $y$, available to the network and signs the block using its own private key PrK, i.e. $x$.

\begin{figure}[htbp]
	\centering{\includegraphics[width=0.65\textwidth]{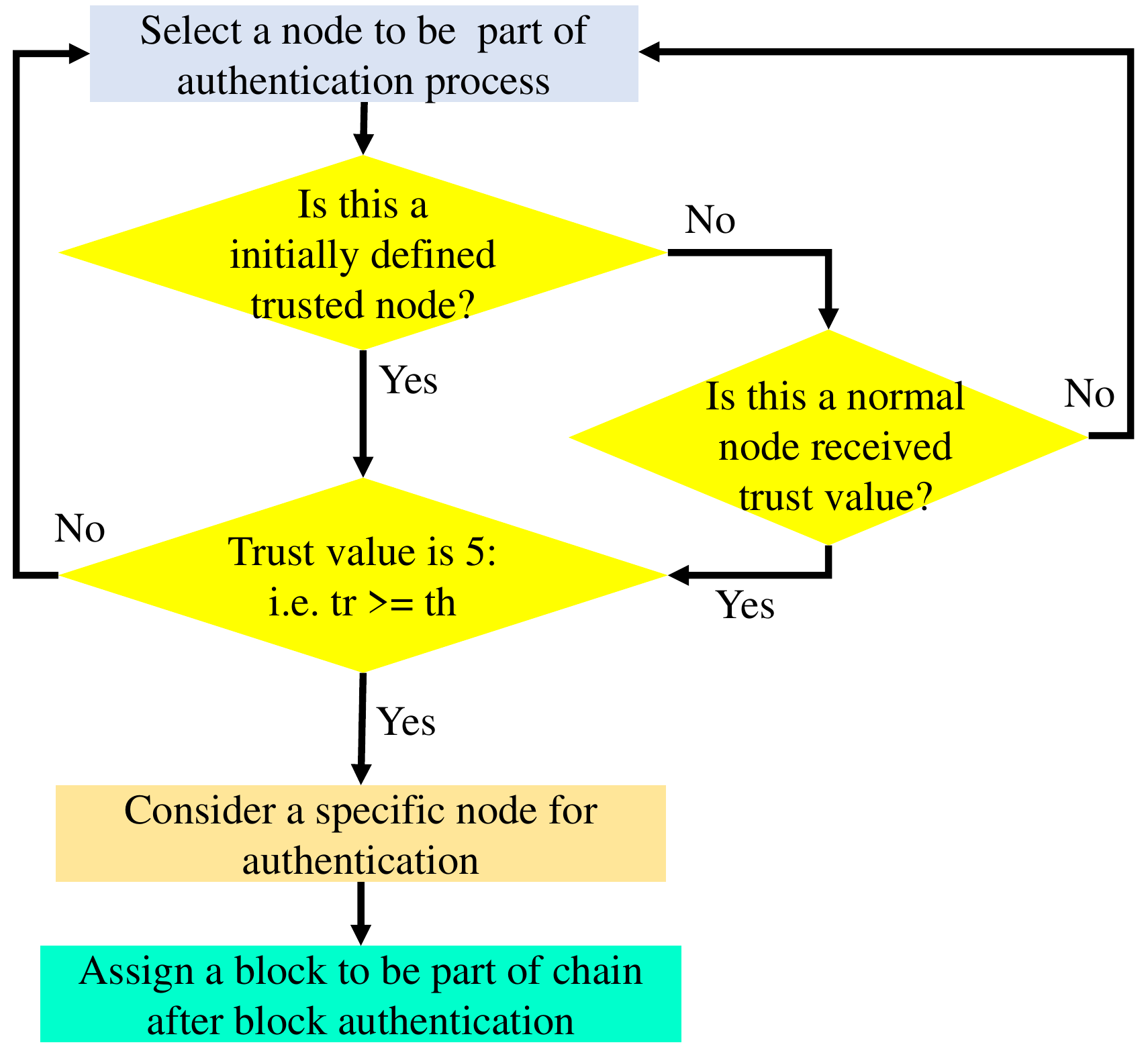}}
	\caption{Steps to select authenticated node for PoAh.}
	\label{fig:PoAh_auth}
\end{figure}

The proposed protocol is specifically designed for resource constrained distributed networks, such as the IoT, where there will be limited number of trusted nodes initialized during the network deployment with a minimum trust value and other network devices with a zero trust value i.e. `tr = 0'. With successful block authentication, an authenticated node’s trust value is increased by 1. Similarly, each fake block authentication will result in decreasing the trust value by `tr = 1'.
\textcolor{black}{After node authentication by the trusted node, other untrusted nodes in the network can also identify the authenticated block to gain trust value i.e. `tr = 0.5'. Importantly, identifying false block authentication can gain high trust value i.e. `tr = 1'.}
With this trusted evaluation process, a trusted node can be out of the process when its trust value is lower than a threshold `tr $<$ th', and a normal node can be part of the authentication process. Here, a threshold `th = 5' is considered and an initial trust value `tr = 10' is assigned to trusted nodes. The procedure for authentication is as shown in Fig. \ref{fig:PoAh_auth}. 

Upon receiving the block for validation, the trusted node finds the source public key, i.e. $y$ for signature verification. This work uses asymmetric cryptography, where the public key is used for signature validation. An attacker cannot extract the private key, i.e. $x$, when all other things are publicly available. This is the discrete log problem. After successful signature verification, the trusted node evaluates the Media Access Control (MAC) address and compares it with the received one for a second round of evaluation. After successful authentication, validated blocks are broadcast by the trusted nodes with PoAh identification. Subsequently, individual users in the network verify the PoAh information to add blocks into the chain. After acceptance of a valid block, the user computes a hash value to link the next block and retrieves the previous block hash value to save into the current block. All the nodes in a network follow this property to maintain the chain as shown in Fig. \ref{fig:PoAh}. The technical steps of the PoAh consensus algorithm are presented in Algorithm \ref{ALG:Proposed_Algorithm}.

\begin{algorithm}[t]
\caption{Procedure of the Proposed PoAh.}
\label{ALG:Proposed_Algorithm}
\SetKwData{Left}{left}\SetKwData{This}{this}\SetKwData{Up}{up}
\SetKwFunction{Union}{Union}\SetKwFunction{FindCompress}{FindCompress}
\SetKwInOut{Input}{Inputs}\SetKwInOut{Output}{Outputs}
\Input{All nodes in the network follow $SHA-256$ hash.
Individual nodes have private $(PrK)$ and public keys $(PuK)$.} 
\BlankLine
\Output{Validated Blocks which are added to the blockchain.}
\BlankLine
 $({Trx}^{+})$ $\rightarrow$ blocks; \tcc{Nodes combine various transactions to form the blocks}
 
 $({S}_{PrK})$(block) $\rightarrow$ broadcast; \tcc{Nodes sign blocks with private key and broadcast to network}

 $({V}_{PuK})$(block) $\rightarrow$ MAC Checking; \tcc{Trusted node verifies signature with source public key}

\If{$Authenticated$}{
$block || PoAH(D)$ $\rightarrow$ broadcast; \tcc{Trusted node broadcasts the authenticated block to network}

$H(block)$ $\rightarrow$ Add blocks into chain; \tcc{If nodes hear from trusted node, they add block to blockchain}
}
\Else{ DROP the block; \tcc{If not authenticated, drop the block}
}
GOTO $(Step-1)$ for next block; 
\end{algorithm}

\section{Analysis of The Proposed PoAh Consensus Algorithm}
\label{Sec:PoAh_Analysis}

%

The Proof-of-Authentication (PoAh) algorithm implements an authentication mechanism in contrast to the validation process used by other consensus algorithms. Authentication uses fewer resources and less energy than other mechanisms, which can be highly advantageous in case of a resource-constrained environment like IoT architectures.

For the implementation of the PoAh algorithm, the block contains the payload data, which can be environmental data collected by the sensors, the identification of the node which is invoking the transaction and the time stamp at which the transaction is initiated. In the current paper, all the devices in the network are connected through a wired or wireless network using a router. Hence the MAC addresses of the devices are used as identification while initiating the transaction. Once a transaction is initiated, the trusted node receives the transactions and starts the validation process. After validating the transaction, the trusted node or the miner adds the block to the local blockchain and retransmits the block to the other nodes in the network. All the other participants add the block only if the block is sent by the trusted node. Once the nodes get blocks from trusted nodes, they add them to the local blockchain ledger in the database. This work makes the following claims for PoAh to validate its scalability, while Section \ref{Sec:Experiments} presents the simulated and testbed evaluations of the PoAh algorithm.

\textbf{\textit{Claim - 1:}} PoAh utilizes minimal resources for block validation.

\textbf{\textit{Proof:}} PoW works as a traditional consensus algorithm for the blockchain, where individual nodes can generate data blocks and miners validate them before being added to the chain. The miner process is nothing but the computation of the inverse of a hash, and it takes the equivalent energy consumed by two households in one day to validate a block \cite{zyskind2015decentralizing}. Utilizing this much energy is not feasible, when it comes to an IoT system. In the IoT, devices are resource constrained with minimal computing power and energy supply. PoAh is proposed to address this issue of evaluating blocks with minimal energy. PoAh introduces an authentication mechanism using a digital signature process. In a cryptographic context, digital signature and hash (not inverse hash) computation is very fast and utilizes minimal energy. This fact is reflected in the experimental results in the following section.   

\textbf{\textit{Claim - 2:}} PoAh requires minimal time compared to PoW without compromising security threats.

\textbf{\textit{Proof:}} From the above claim, it is found that PoW consumes significant energy while validating a block. In addition to this, PoW takes approximately 10 minutes to evaluate a block \cite{zyskind2015decentralizing}. This is not acceptable in any kind of IoT application. IoT applications are mainly deployed for real-time monitoring purposes and 10 minutes to evaluate a block is not acceptable under any circumstances. 
PoAh proposes to address this issue to evaluate blocks in minimal time. As PoAh utilizes authentication to validate blocks, cryptographic authentication takes significantly less time. From experimental evaluations, it is found that PoAh taking time in seconds, i.e. it is 1000 times faster. \textcolor{black}{Integration of cryptographic authentication and digital signatures ensure the security level of PoAh \cite{dorri2017blockchain}}. Hence, it can be concluded that PoAh is ideally suited for IoT applications. 

\textbf{\textit{Claim - 3:}} PoAh provides substantial security while integrating a blockchain based decentralized security solution to the IoT.

\textbf{\textit{Proof:}} By introducing PoAh as a consensus algorithm for the blockchain in IoT, the proposed decentralized security solution provides sustainable security infrastructure. The IoT does not requires the same level of security as required for cryptocurrencies \cite{puthal2018blockchain}. IoT applications require real-time security with proper authentication of data and source, wherein a cryptographic solution is sufficient protection.  
PoAh integrates with the existing cryptographic concept of PoW but ignores the block evaluation of computing the inverse of a hash.  With the SHA-256 hash (from Algorithm – 1) and digital signatures, the proposed PoAh provides a decentralized security solution by providing the same level of security as asymmetric cryptography. 

\textcolor{black}{Furthermore, we are addressing two major weaknesses of current blockchain consensus, namely unstable network connectivity (which may prevent all peers from communicating), and the 51\% attack in the network. In PoAh, all the network devices are eligible to generate blocks, whereas only trusted nodes are authenticating them. In any unfavorable situation arising from these weaknesses,  are not broadcast to all peers. Only reachable (solves the unstable network problem) and already trusted peers (solves the 51\% attack issue) can authenticate and add blocks into the chain. As a result, the 51\% attack weakness of PoW is addressed due to the dynamic nature of trust values.}

\textcolor{black}{\textbf{\textit{Claim - 4:}} PoAh provides a better platform for IoT compared to other blockchain consensus algorithms.}

\textcolor{black}{\textbf{\textit{Proof:}} The consensus algorithm is the backbone of the blockchain and makes the network and the process decentralized. However, current consensus algorithms widely used, such as Proof of Work (PoW), Practical Byzantine Fault Tolerance (PBFT), Proof of Authority (PoAu), and Proof of Elapsed Time (PoET) are very expensive in terms of security and resource requirements \cite{andoni2019blockchain}. PoW requires approximately 10 minutes to valid a block and approximately 1 hour for the block to become accepted in the chain. PBFT requires at least 2/3 of the network devices to behave honestly by signing transactions and message overhead may increase significantly as the size of the network increases, affecting both speed and scalability. PoAu consensus represents a more centralized approach most appropriate for governing or regulatory bodies, and is currently also proving popular with utility companies in the energy sector. PoET is developed by Intel, meaning that trust is still required towards a single authority  \cite{hertig2016intel}.}

\textcolor{black}{PoAh is developed to address the above consensus drawbacks for scalable IoT deployment.}

\section{Experimental Evaluations}
\label{Sec:Experiments}

\textcolor{black}{The proposed PoAh consensus algorithm has been validated using a simulation environment (for large scale study), as well as a hardware based experimental test-bed for real-life scenarios. The details of both options, including the experimental setup to validate the proposed model's sustainability, and results analysis are presented in this Section.}

\subsection{Simulation Evaluation}

\begin{figure}[htbp]
   \centering{\includegraphics[width=0.60\textwidth]{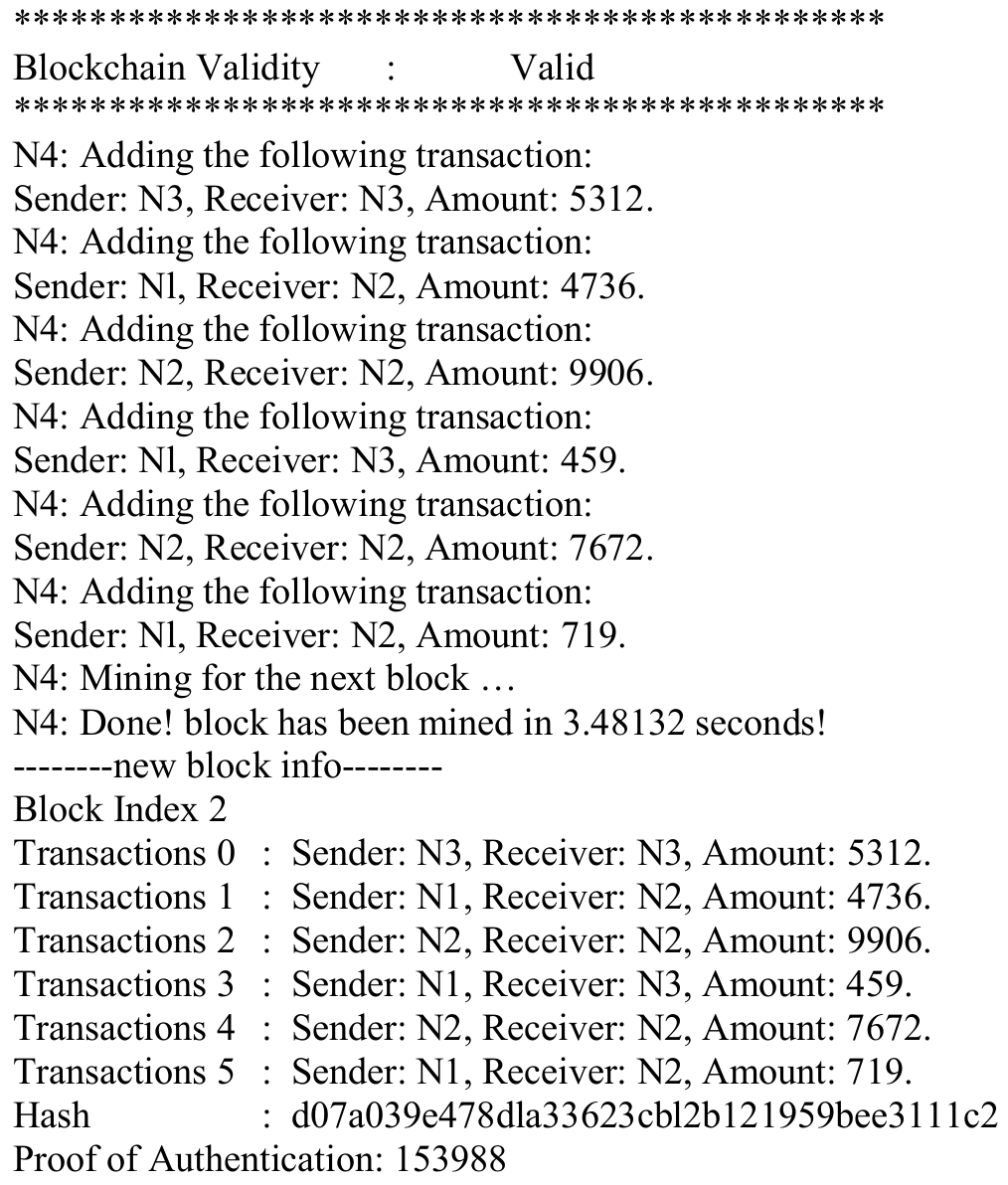}}
   \caption{Sample output of PoAh.}
   \label{fig:PoAh-out}
\end{figure}

The proposed PoAh algorithm is simulated to evaluate its sustainability in the Python programming language. In this experimental setup, an environment of five nodes in the network has been used, where two nodes are trusted nodes to work as miners and 35 bytes are taken as the size of block. Multiple transactions are incorporated into a block before being broadcast. This work uses the ElGamal cryptographic system for signature, verification, MAC and encryption. In the simulation, the block format is created as $<$Source ID, ``Signature'', MAC, Trx1, Trx2, $…>$ for PoAh testing. 
The output screenshot of the PoAh evaluation is shown in Fig. \ref{fig:PoAh-out}.


\begin{figure*}[t]
	\centering
\includegraphics[width=0.992\textwidth]{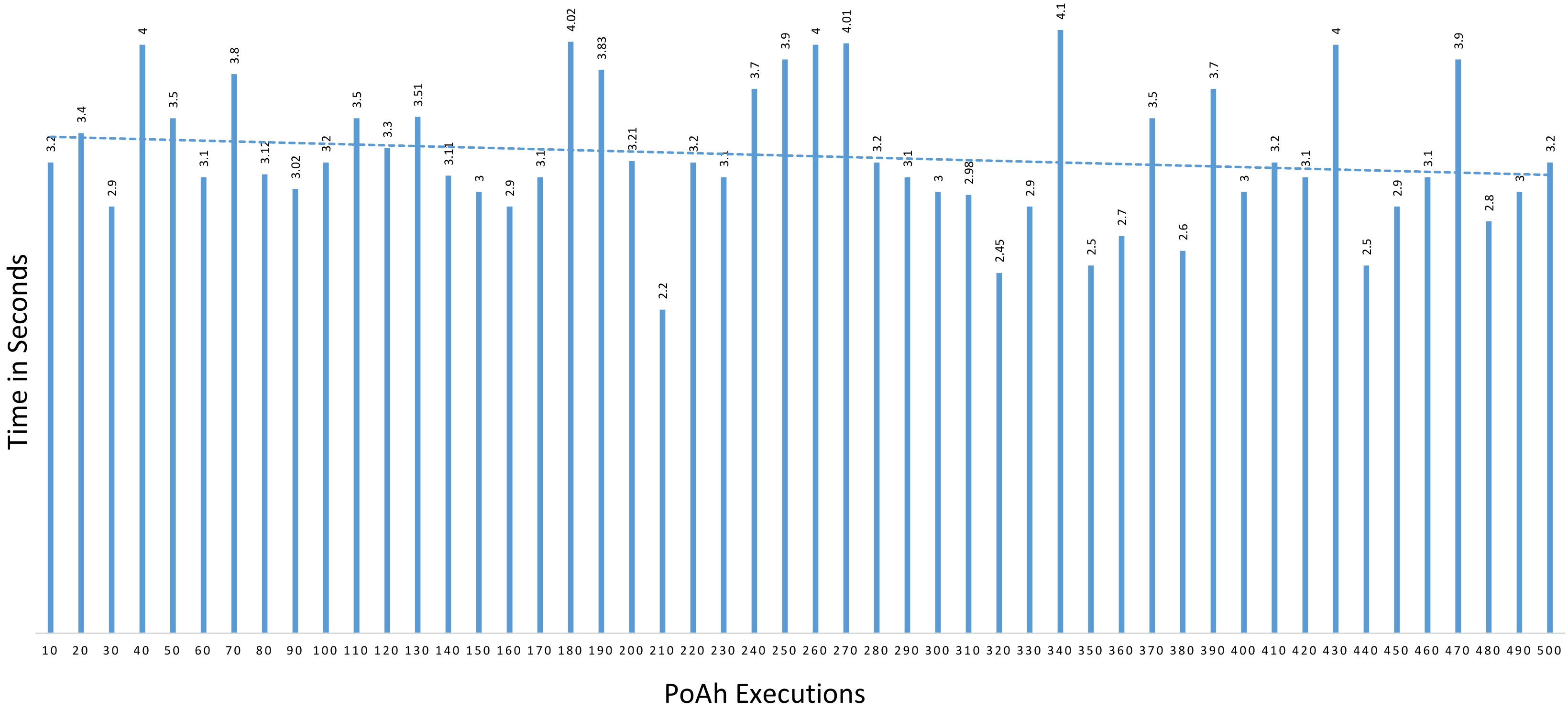}
	\caption{\textcolor{black}{Time taken by PoAh for authentication of blocks.}}
	\label{FIG:PoahSimulation}
\end{figure*}

The network users generate blocks and sign them using own their private key and broadcast them for validation. Trusted nodes use the source node public key to validate blocks. If blocks are authenticated successfully, the trusted node broadcasts the blocks again to add into the chain. In this simulation environment, results from 500 iterations show that the average time for PoAh is 3.34 seconds. The 500 iteration results from the simulation are listed in Fig. \ref{FIG:PoahSimulation}, where they are computed from the Algorithm \ref{ALG:Proposed_Algorithm} steps. 

\subsection{Testbed Evaluation }

\textcolor{black}{The Proof-of-Authentication (PoAh) consensus algorithm was implemented in a real-time hardware testbed for evaluation of performance. The testbed included 6 Raspberry Pi single board computers, where five boards were collecting the environmental data using various sensors. Among the five Raspberry Pis, three of them were Raspberry Pi 1 Model B+ and the other two were Raspberry Pi 3 Model B. This was considered to evaluate the performance of the consensus algorithm when different processing powers were utilized. The Raspberry Pi 1 Model B+ is equipped with a 700 MHz Cortex single core processor with 512 MB of RAM whereas the Raspberry Pi 3 Model B is equipped with a Quad Core 1.2 GHz ARM Cortex - A53 CPU and 1 GB LPDDR2 RAM. These Raspberry Pi boards were used for collecting the data and for validating it. In addition, a trusted node was deployed in the network. As a trusted node, a Raspberry Pi 3 Model B+ was deployed which is equipped with a Quad Core 1.4 GHz  ARM Cortex A53 CPU and 1GB of LPDDR2 RAM. The trusted module also has the capability of connecting to a 5 GHz wireless network which gave it the advantage of better and faster connectivity to the network.} \textcolor{black}{This testbed is of sufficient complexity to allow real-time evaluation of the PoAh algorithm and is not intended as a commercial deployment demonstration.}

\textcolor{black}{For the experimental setup, the single-board computers that are connected using Ethernet cables are the Raspberry Pi} 1 Model Bs as they do not have wireless communication capabilities. All the devices are connected through a router and are connected using wired and wireless connectivity wherever possible. Once all the devices are connected to each other, the PoAh algorithm will start authenticating the transactions performed by the nodes. The load of the transactions is the time stamp, the environmental data collected and the identity of the node that is collecting the data. Every participant in the network stores the blockchain ledger locally using an SQLite database. For real-time evaluation, 300 transactions were performed by all the nodes and various parameters were observed for each transaction as discussed in the following subsections.

\begin{figure}[htbp]
	\centering{\includegraphics[width=0.65\textwidth]{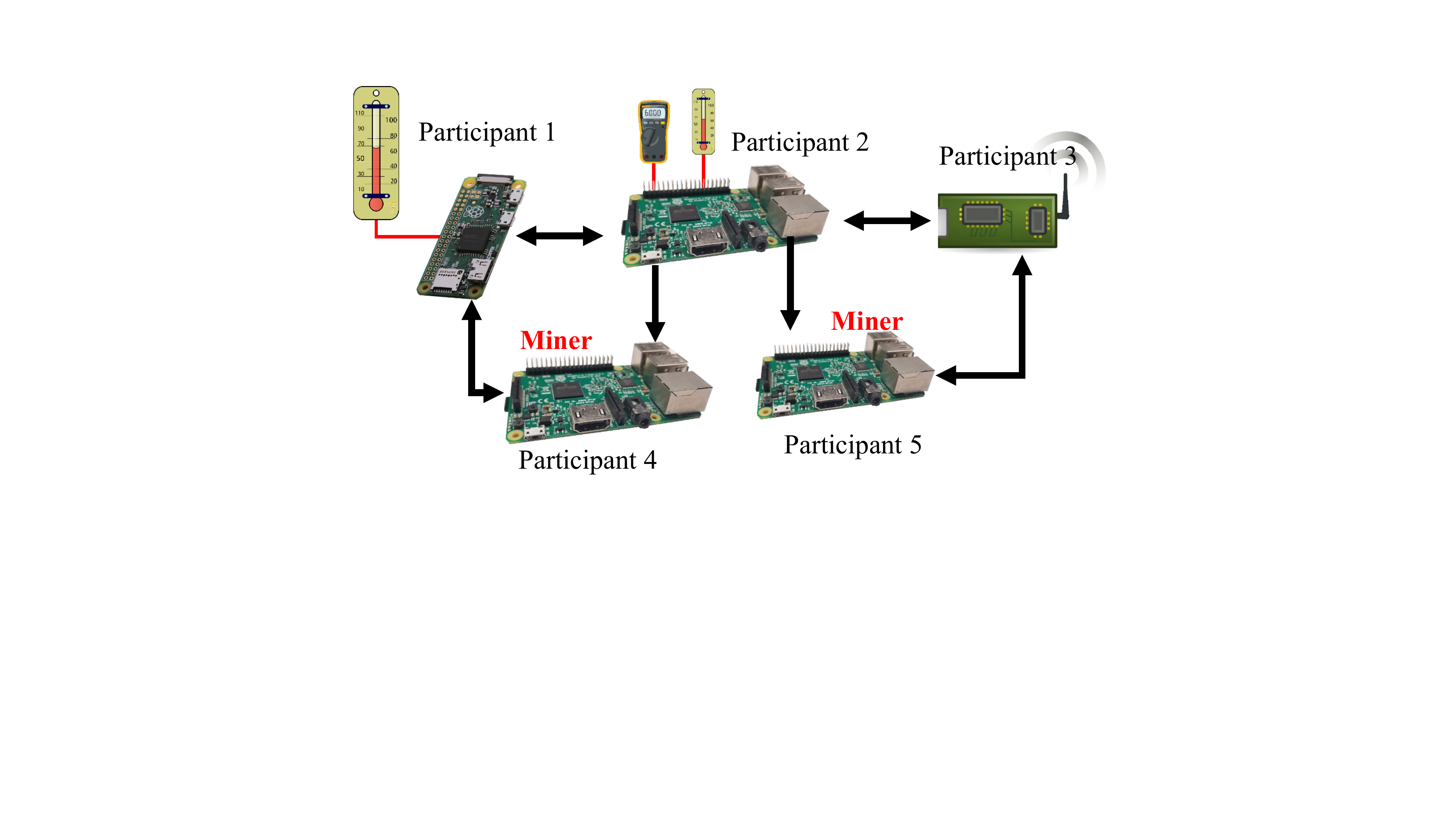}}
	\caption{Block processing with transactions and hashing.}
	\label{fig:Validation_Testbed}
\end{figure}
\begin{figure*}[htbp]
	\centering{\includegraphics[width=0.99\textwidth]{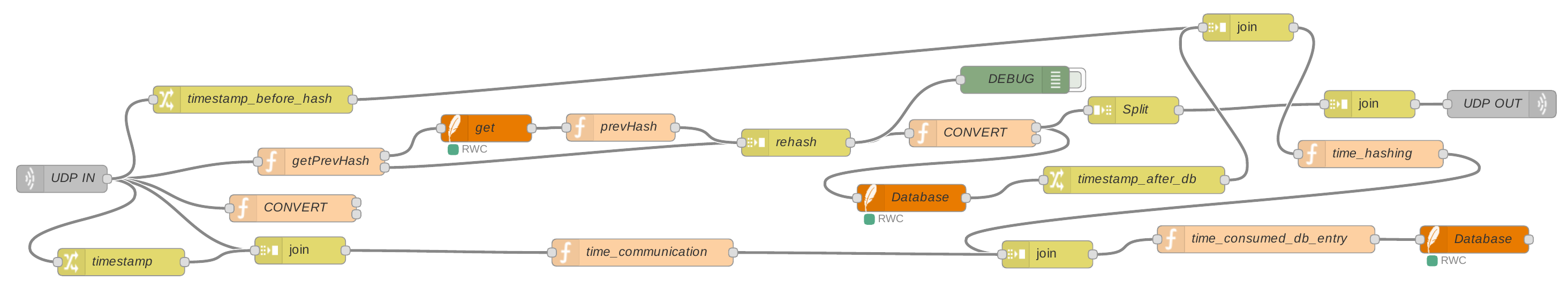}}
	\caption{Node-RED features in PoAh experiment.}
	\label{fig:Node-RED}
\end{figure*}

The Node-RED development tool is used for prototyping the PoAh blockchain consensus algorithm on the Raspberry Pi. Node-RED is a flow-based programming tool used for the development of IoT applications. The interface can be accessed through a web browser which makes it optimal for a distributed work environment like the implementation of the PoAh consensus algorithm. The server and the client node software was developed and deployed using the Node-RED interface on the Raspberry Pi single board computers. The Node-RED development environment is not confined by the operating system, which makes the development of PoAh deployable on various environments and not restricted to the Raspberry Pi. Fig. \ref{fig:Validation_Testbed} shows the implementation of the PoAh algorithm on a Raspberry Pi Mode 3B+. As shown in the figure, once the data is received from the network, the hash of the previous block is fetched from the blockchain and is sent cascading it to the data received. After authentication of the device that sent the data, the server then calculates the hash of the entire block which contains the data received and the previous hash. This is then added to the blockchain at the server and transmitted to the other nodes to be added to their respective blockchains. The details of the Node-RED program are shown in Fig. \ref{fig:Node-RED}.

\textcolor{black}{It needs to be noted that we tried to run PoW and PoS in the real-life resource-constrained IoT devices available in our testbed. We observed that even one mining process could not be completed for these heavy-duty algorithms. We observed that the testbed hardware platform doesn't complete even after running it for days. This indicates that it is impossible to run these heavy-duty consensus algorithms in resource-constrained IoT devices. If we run these IoT devices under the assumption that they are deployed in possibly remote locations in real-life applications, like smart cities \cite{Mohanty_CEM_2016-Jul} and powered by batteries, then obviously the battery won't last enough to even complete one mining process for these heavy-duty algorithms.}

\subsection{Transaction Time}

\begin{figure}[t]
	\centering{\includegraphics[width=0.60\textwidth]{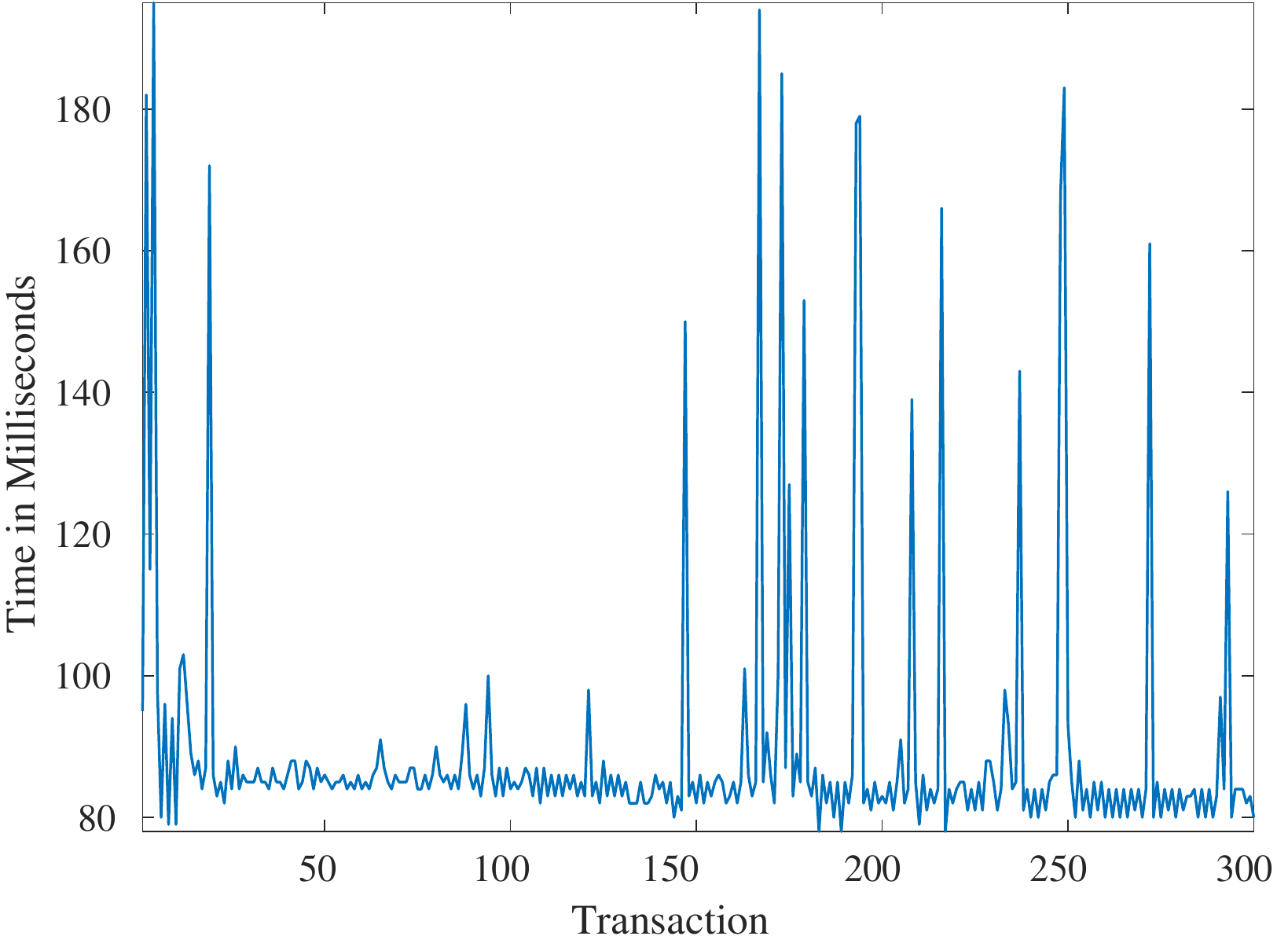}}
	\caption{Time taken for the transaction to reach validator from the ClearPi.}
	\label{fig:CommunicationToBlackPi2}
\end{figure}

\begin{figure}[t]
	\centering{\includegraphics[width=0.60\textwidth]{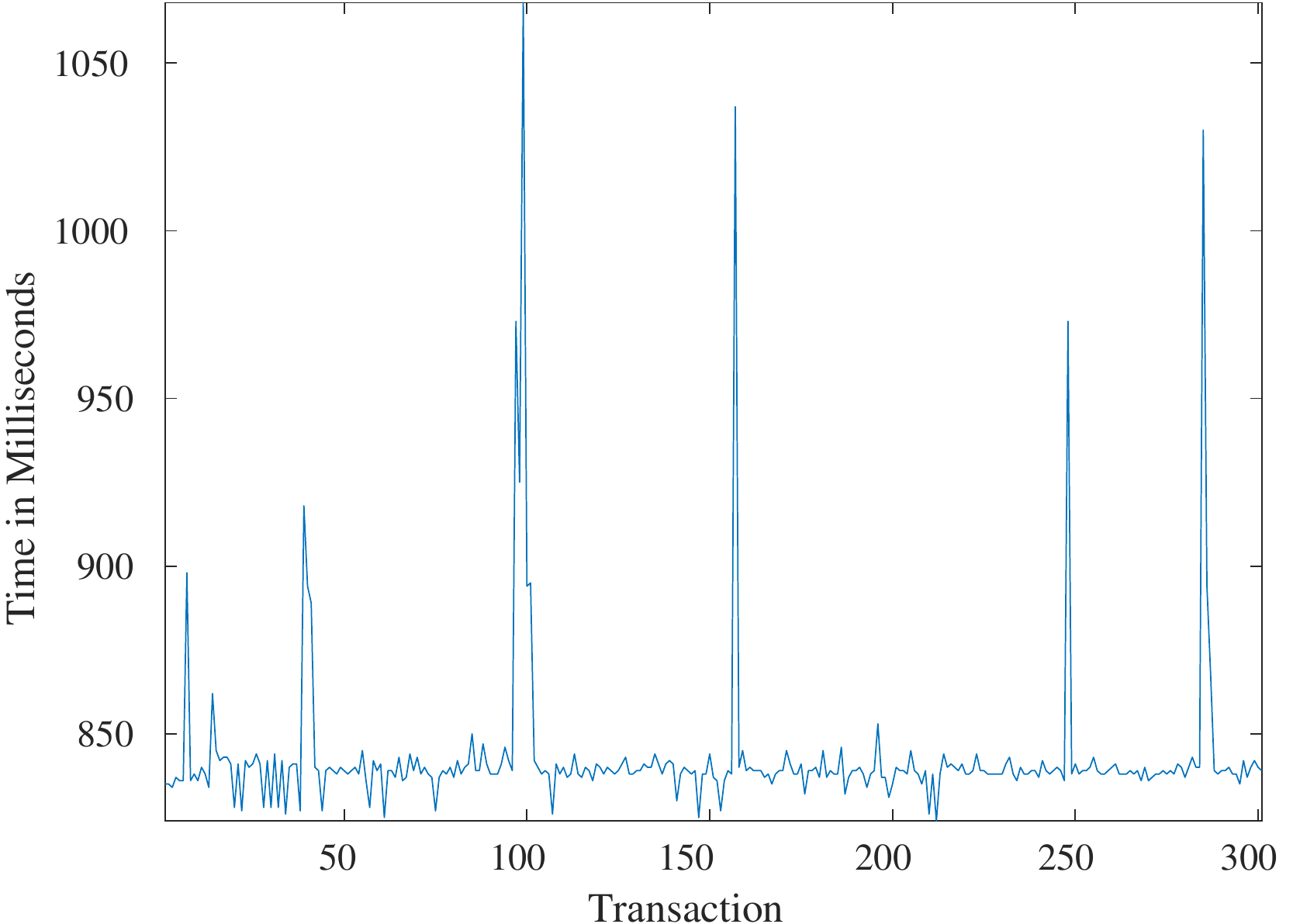}}
\caption{Time taken for the block to be validated and to be added to the blockchain at the BlackPi.}
	\label{fig:Add_Block_To_Blockchain}
\end{figure}

\begin{figure}[htbp]
	\centering{\includegraphics[width=0.60\textwidth]{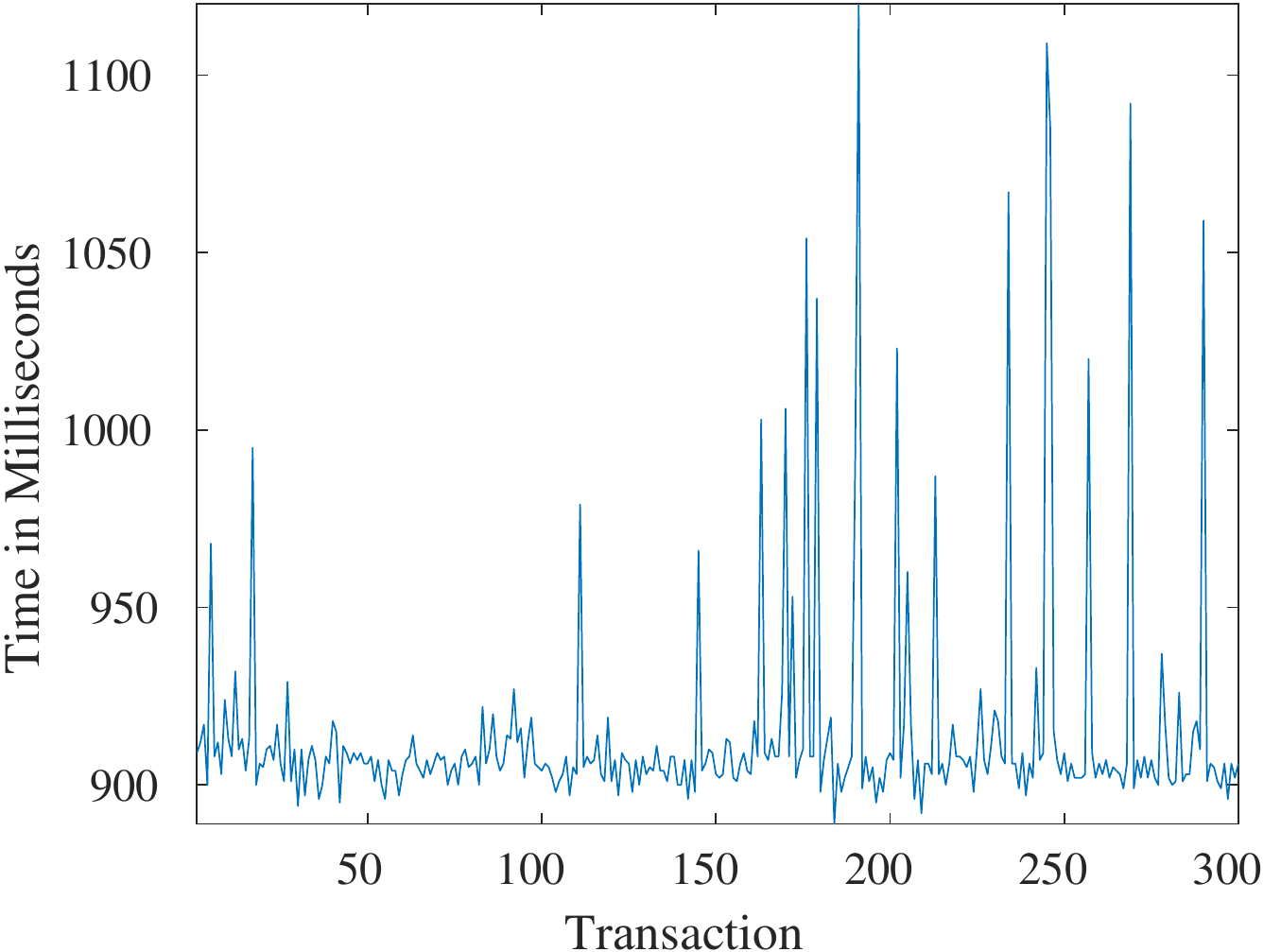}}
	\caption{Time taken for the transaction to reach ClearPi from block creation and after validation.}
	\label{fig:Comm_Add_Block_Rpi2_ClearPi}
\end{figure}

Fig. \ref{fig:CommunicationToBlackPi2} shows the time taken by a transaction to reach the miner or trusted node. The transactions have taken around 100ms to 200ms and this depends on the connection type that was used for the transaction. Fig. \ref{fig:Add_Block_To_Blockchain} presents the time taken by the miner or the trusted node to authenticate the data sent and add to the blockchain at the validation node locally. Once the validation is complete, the trusted node will broadcast the transaction back to the network nodes at this point. Once the other nodes obtain the transaction from the trusted node or the miner, they add the transaction to their local blockchain. Once this is done, a transaction is considered complete. The time taken for the transaction to be completed is presented in Fig. \ref{fig:Comm_Add_Block_Rpi2_ClearPi}.

\begin{figure}[htbp]
	\centering
\subfloat[For a Raspberry Pi Model 2B (a Clear Pi)]{\label{fig:Add_Block_Rpi2_ClearPi}\includegraphics[width=0.45\textwidth]{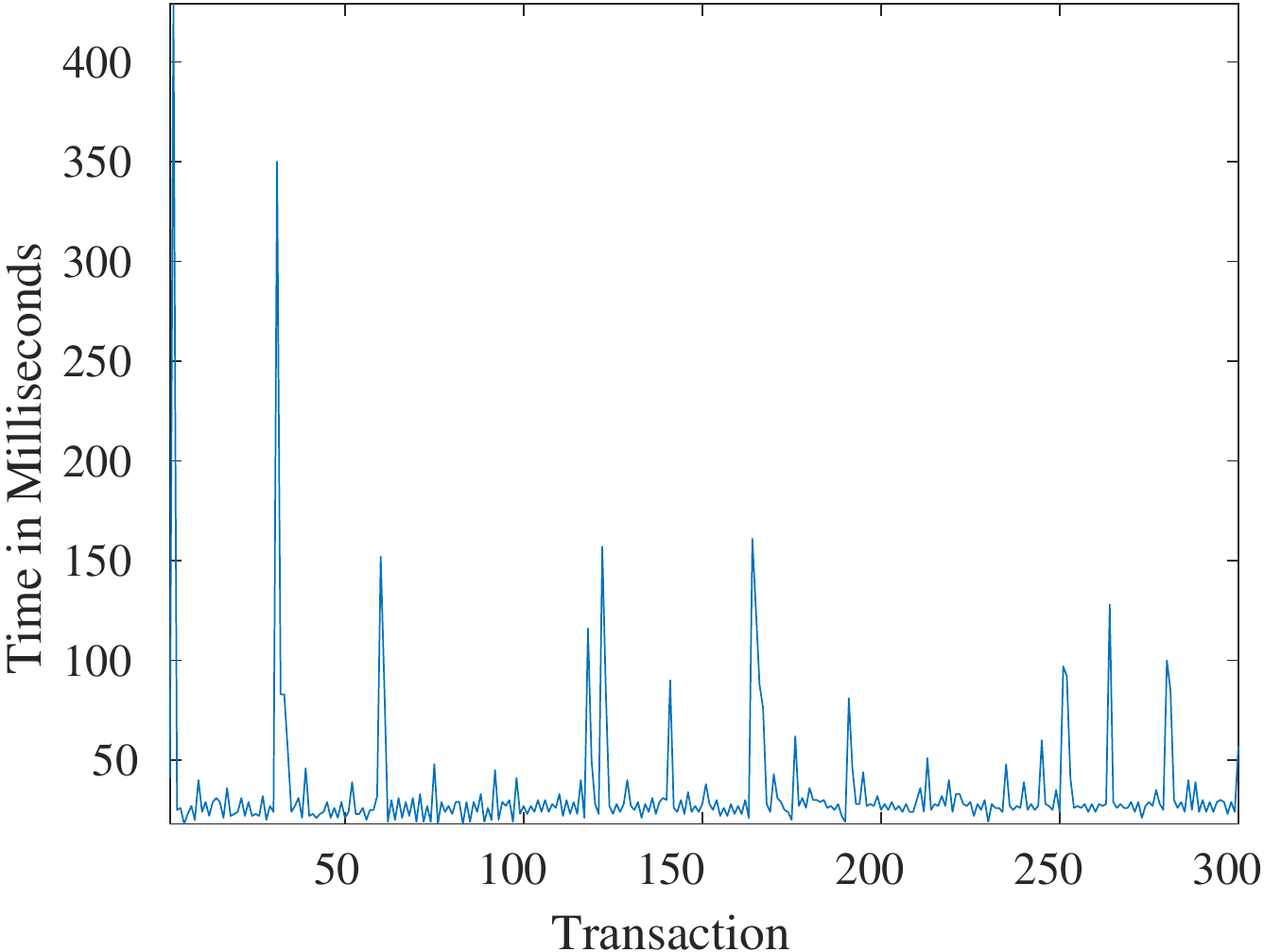}}\quad
\subfloat[For Raspberry Pi Model 1 (a Clear Pi)]{\label{fig:Add_Block_Rpi1_ClearPi}\includegraphics[width=0.45\textwidth]{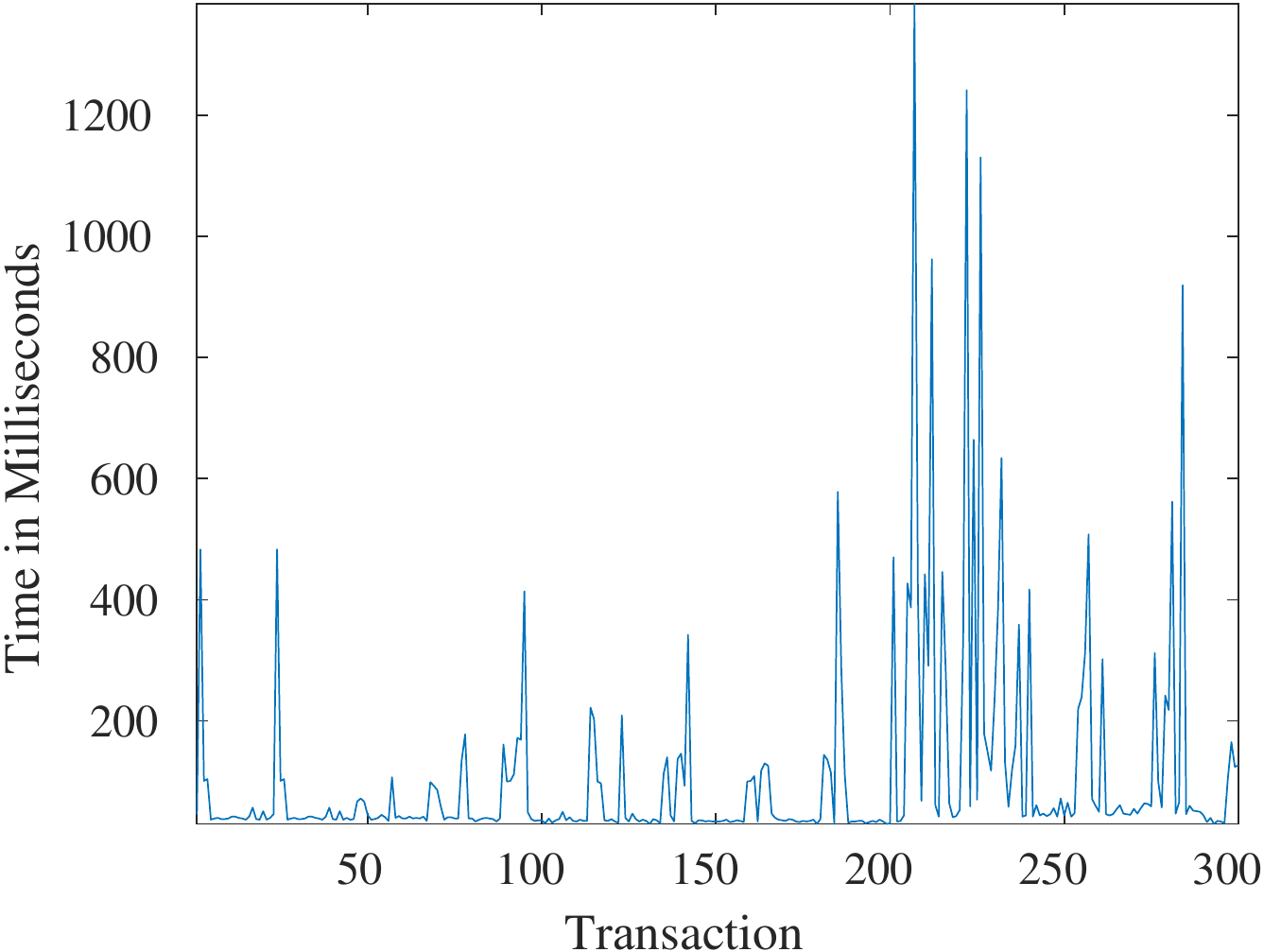}}
\caption{Time taken to check if Black Pi sent the transaction and add it to the Blockchain (Spikes when traffic is high).}
	\label{FIG:Time_Taken_by_Raspberry-Pi}
\end{figure}

Fig. \ref{fig:Add_Block_Rpi2_ClearPi} and Fig. \ref{fig:Add_Block_Rpi1_ClearPi} show the time required for 300 transactions to be added by different Raspberry Pis. Fig. \ref{fig:Add_Block_Rpi2_ClearPi} shows the data pertaining to the Raspberry Pi 2 Model B+. Once the Raspberry Pi 2 Model B+ receives the data from a miner or validator node, the time consumed by that single board computer to check the identity of the sender and add to the blockchain is shown in Fig. \ref{fig:Add_Block_Rpi2_ClearPi}. Similarly, in the case of a Raspberry Pi 1 Model B, the time taken by the single board computer to check the received data and add it to the blockchain is presented in Fig. \ref{fig:Add_Block_Rpi1_ClearPi}.



As shown in Fig. \ref{fig:Node-RED}, various timestamps were generated during the process for calculating the time consumed for various processes. At the server node, a timestamp $t_{sr}$ is generated when the message is received at the node. When the message traverses through various functions in NodeRed, timestamp $t_{sh}$ is generated at the server and timestamps $t_{cr}$ and $t_{cv}$ are generated at the client. The time required for authentication by the server is:
\begin{equation}
	\delta t_{sa} = t_{sh} - t_{sr},
\end{equation}
where $t_{sh}$ is the timestamp after validation and hashing are complete at the server.

The time required by the client to check if the message is sent by the server and add it to the blockchain is:
\begin{equation}
	\delta t_{ca} = t_{ch} - t_{cr},
\end{equation}
where $t_{cr}$ is the timestamp generated when the authenticated message is received at the client and $t_{ch}$ is the timestamp generated when the message is added to the blockchain at the client node. The mean and standard deviations presented in Table \ref{TBL:Adding_Blocks_To_Blockcahin} and Table \ref{TBL:Communication_Between_Devices} are using $\delta t_{sa}$ and $\delta t_{sa}$ calculated over a period of 300 transactions.

The timestamps $t_{sr}$, $t_{sh}$, and $t_{cr}$ are also used for calculating the time required for communication between the devices during the transactions. This helps in finding the latency and the overhead added by different communication mechanisms. When the sensor data is collected by the client node and a transaction is initiated, an initial timestamp is added to the block, $t_{i}$. The time taken to complete one transaction is the following:
\begin{equation}
	\delta t_{tx} = t_{ch} - t_i,
\end{equation}
where $t_{ch}$ is the corresponding client node timestamp where the transaction is added to the blockchain.

\subsection{Power Consumption}

\begin{figure*}[htbp]
	\centering
	\subfloat[Maximum power consumption of Raspberry Pi 3 Model B+]{\label{fig:Power_Consumption_Rpi3}\includegraphics[width=0.28\textwidth]{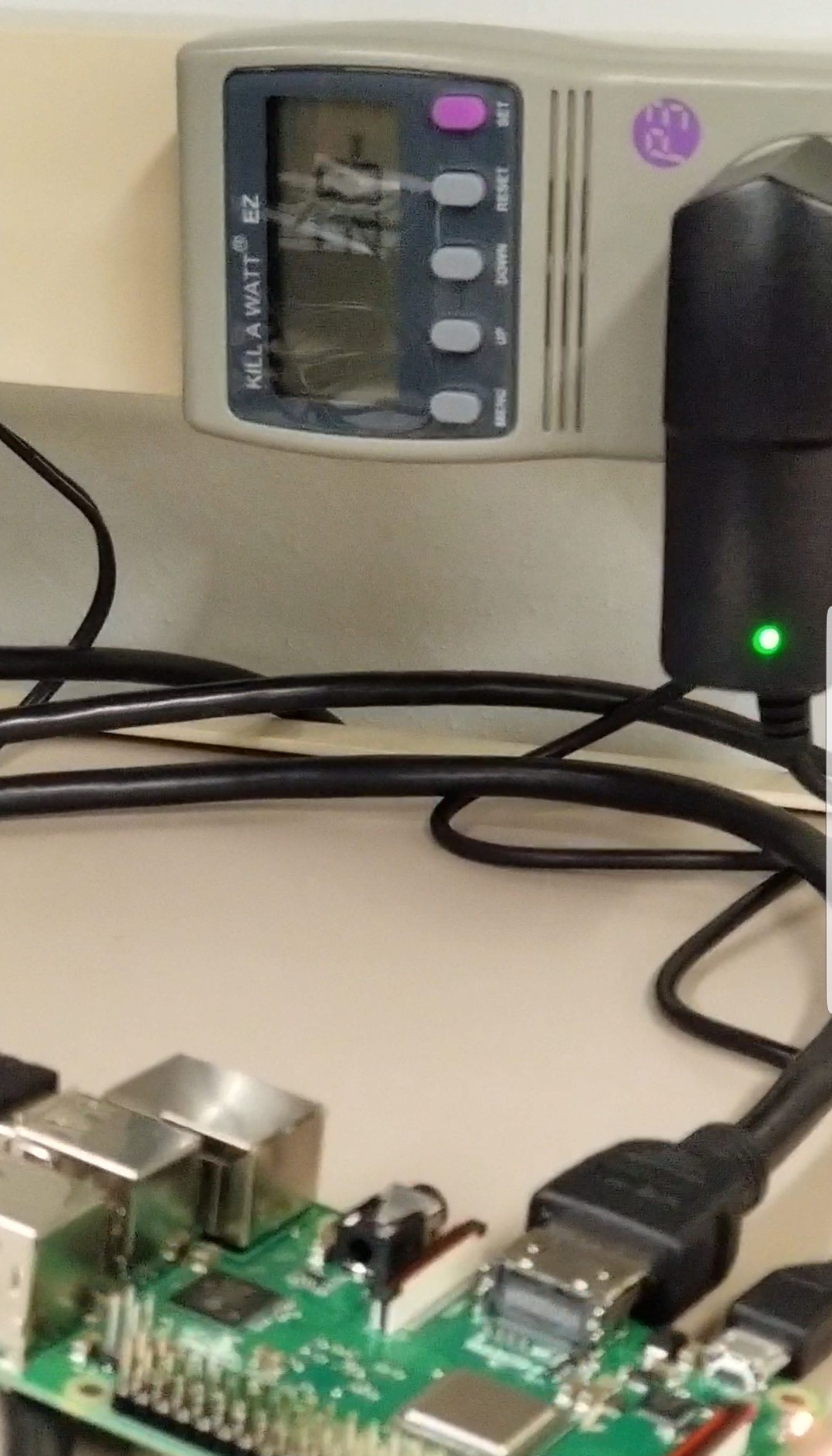}}\quad
	\subfloat[Maximum power consumption of Raspberry Pi 2 Model B+]{\label{fig:Power_Consumption_Rpi2}\includegraphics[width=0.27\textwidth]{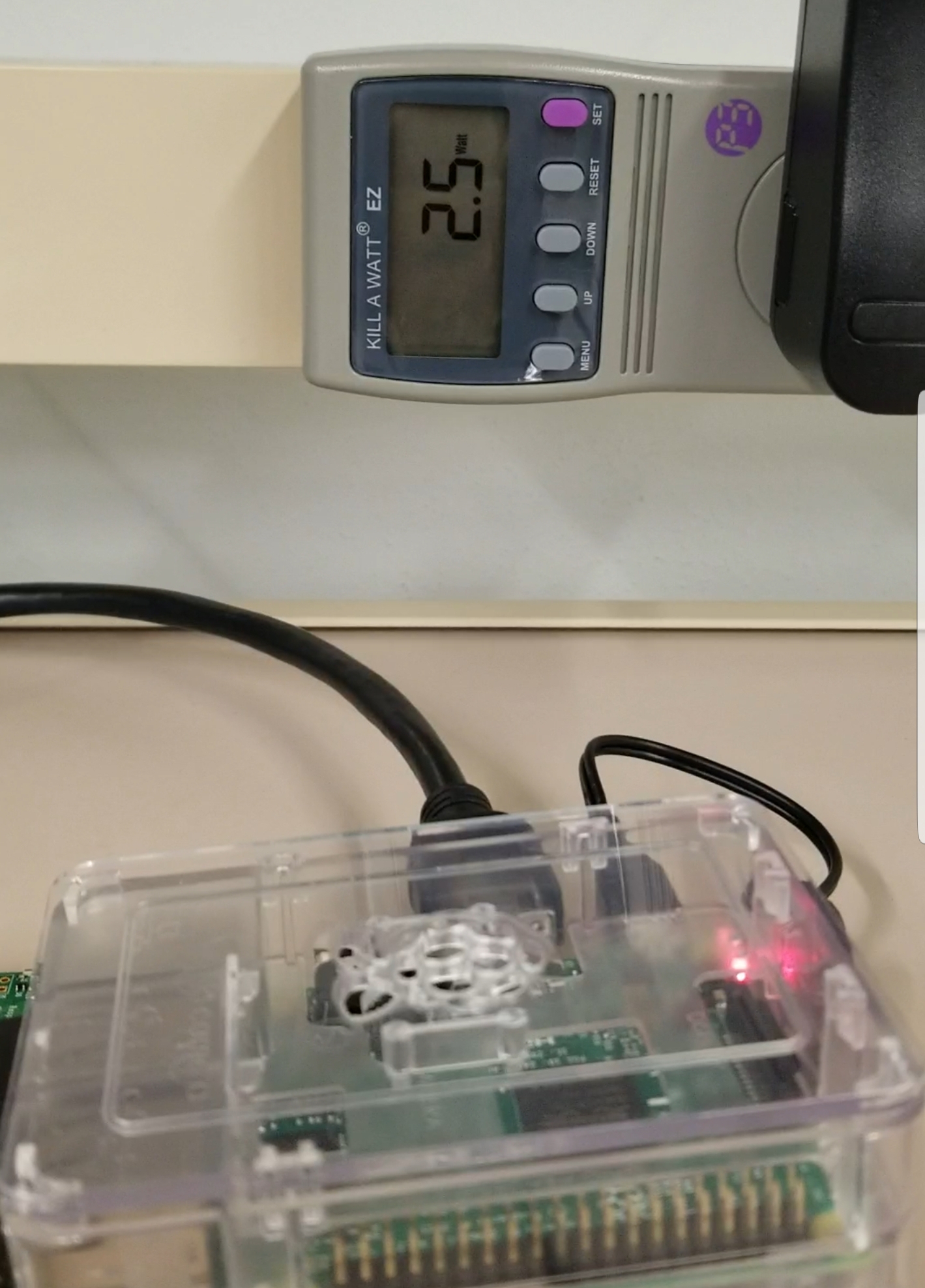}}\quad
	\subfloat[Maximum power consumption of Raspberry Pi 1 Model B+]{\label{fig:Power_Consumption_Rpi1}\includegraphics[width=0.26\textwidth]{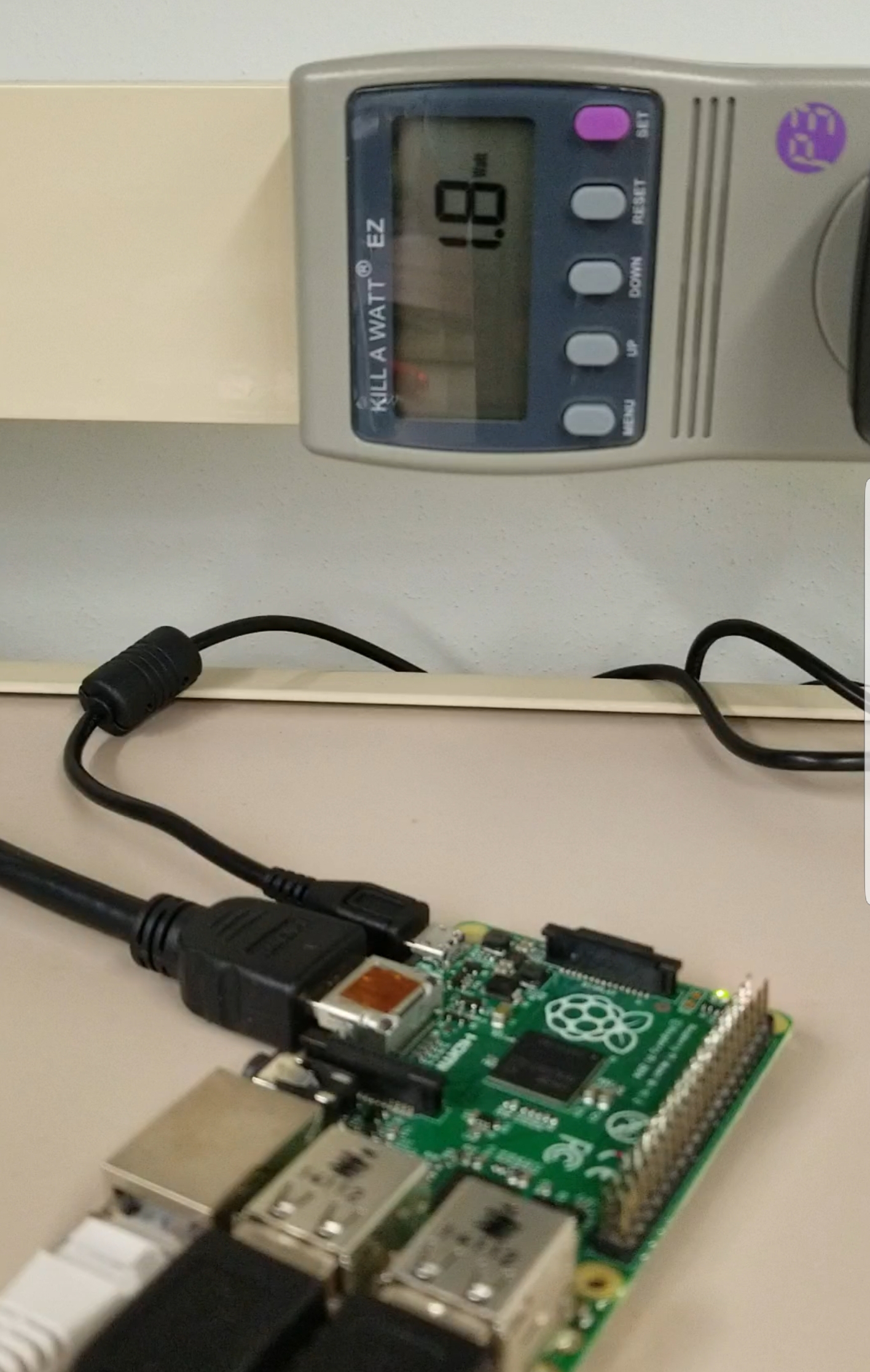}}
	\caption{Experimental setup for power consumption measurement of different versions of Raspberry Pi.}
	\label{FIG:Experimental_Setup_Power_Consumption}
\end{figure*}

Besides the time consumed and the processing power of the entire system, another major challenge in the case of a blockchain network is the power consumption. Fig. \ref{FIG:Experimental_Setup_Power_Consumption} shows the experimental setup for measuring the power consumption of the system. \textcolor{black}{The minimum and maximum power consumption of different Raspberry Pis is listed in Table \ref{TBL:POWER_CONSUMPTION}.} An electrical meter is used to measure the power consumption of the Raspberry Pis as shown in the figure. It is connected to the power outlet and the Raspberry Pi board power supply is connected through the electrical meter. Two measurements were considered during the experiment. Once when the system is idle and once when the Raspberry Pi is processing or adding the block to the blockchain.

\begin{figure}[htbp]
	\centering{\includegraphics[width=0.90\textwidth]{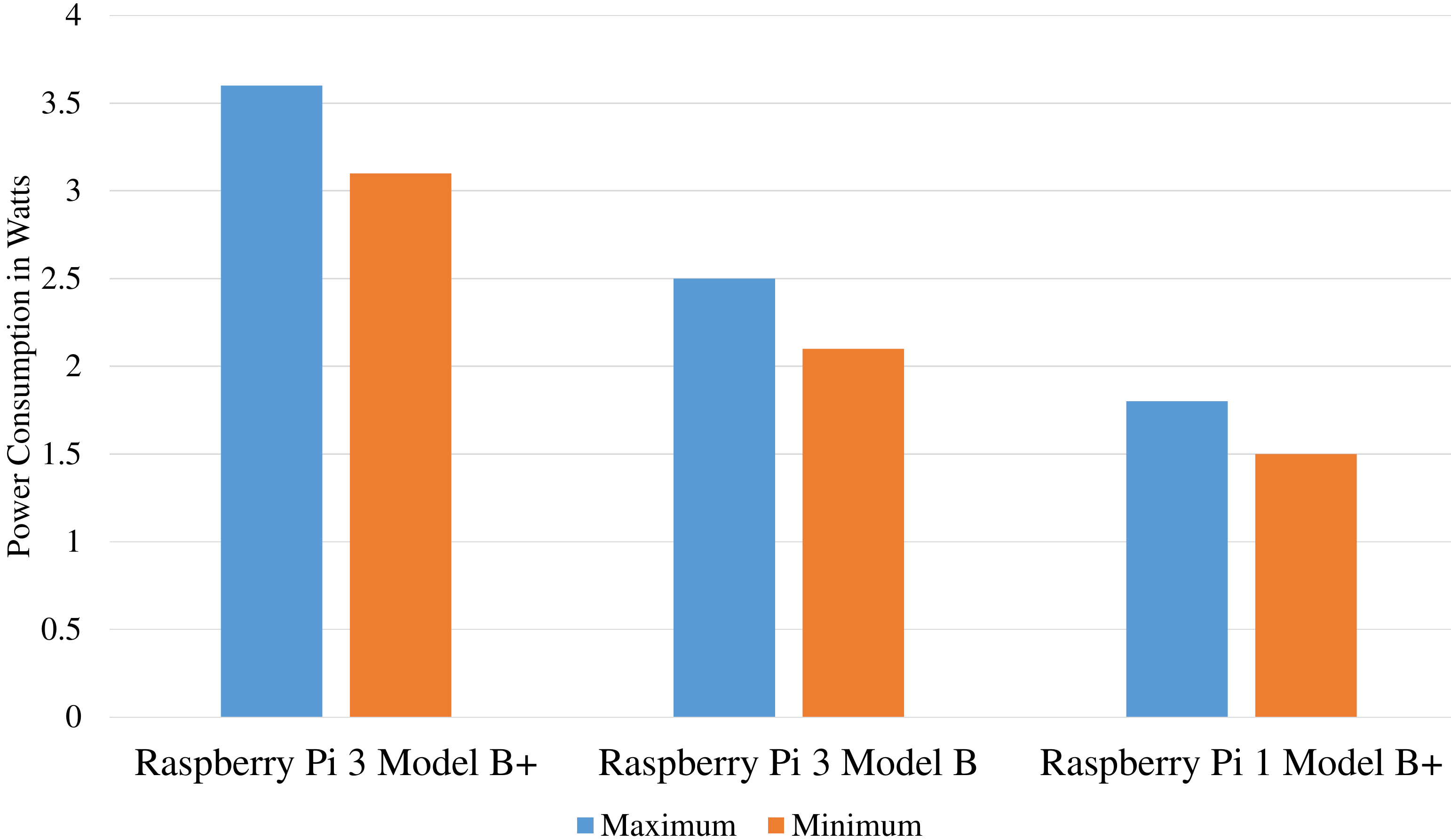}}
	\caption{Power consumption of different Raspberry Pi versions while running the PoAh blockchain.}
	\label{fig:Power_Consumption}
\end{figure}

\begin{figure*}[t]
	\centering
	\subfloat[BlackPi]{\includegraphics[width=0.45\textwidth]{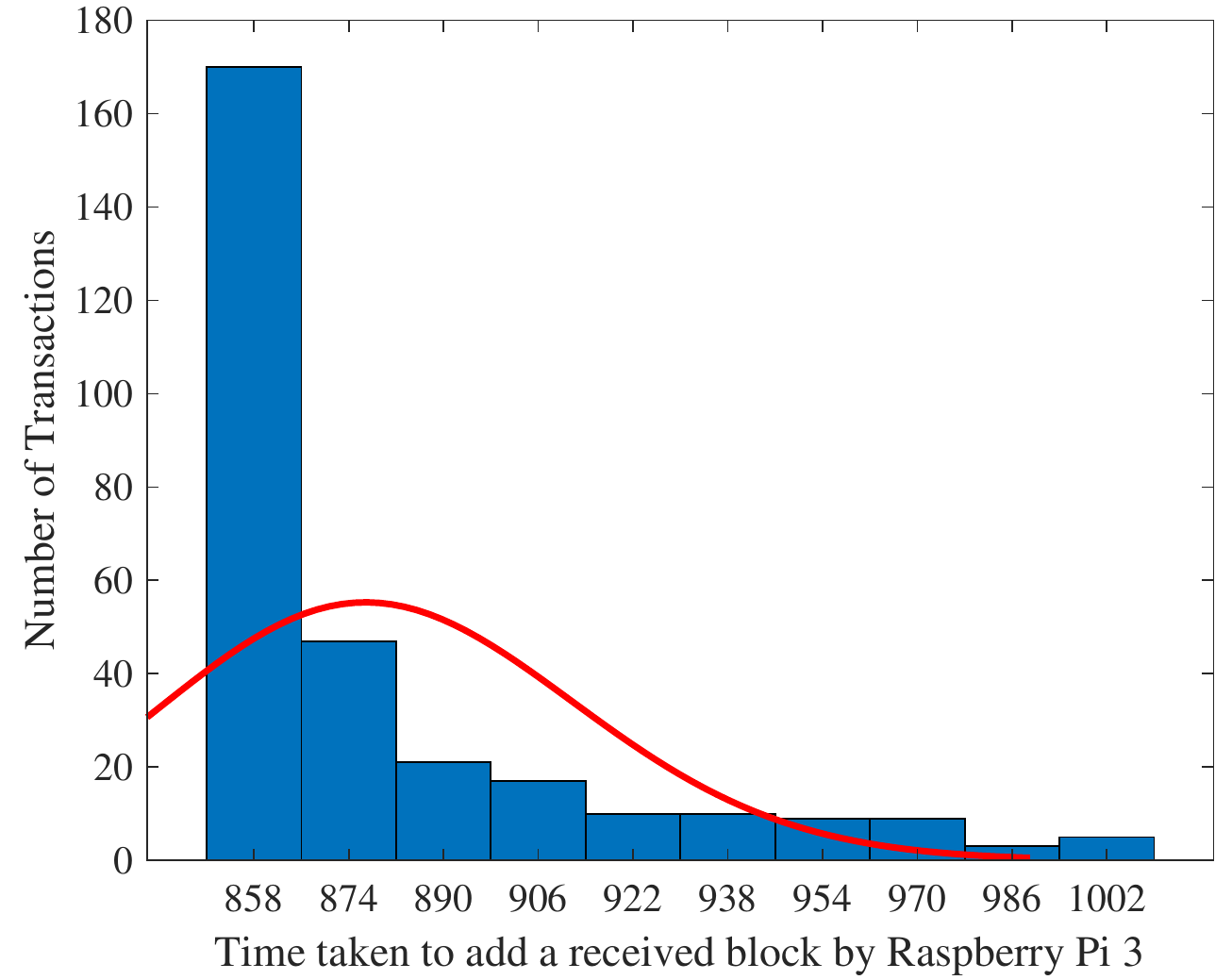}}\qquad
	\subfloat[ClearPi (Raspbery Pi 2 Model B+)]{\includegraphics[width=0.45\textwidth]{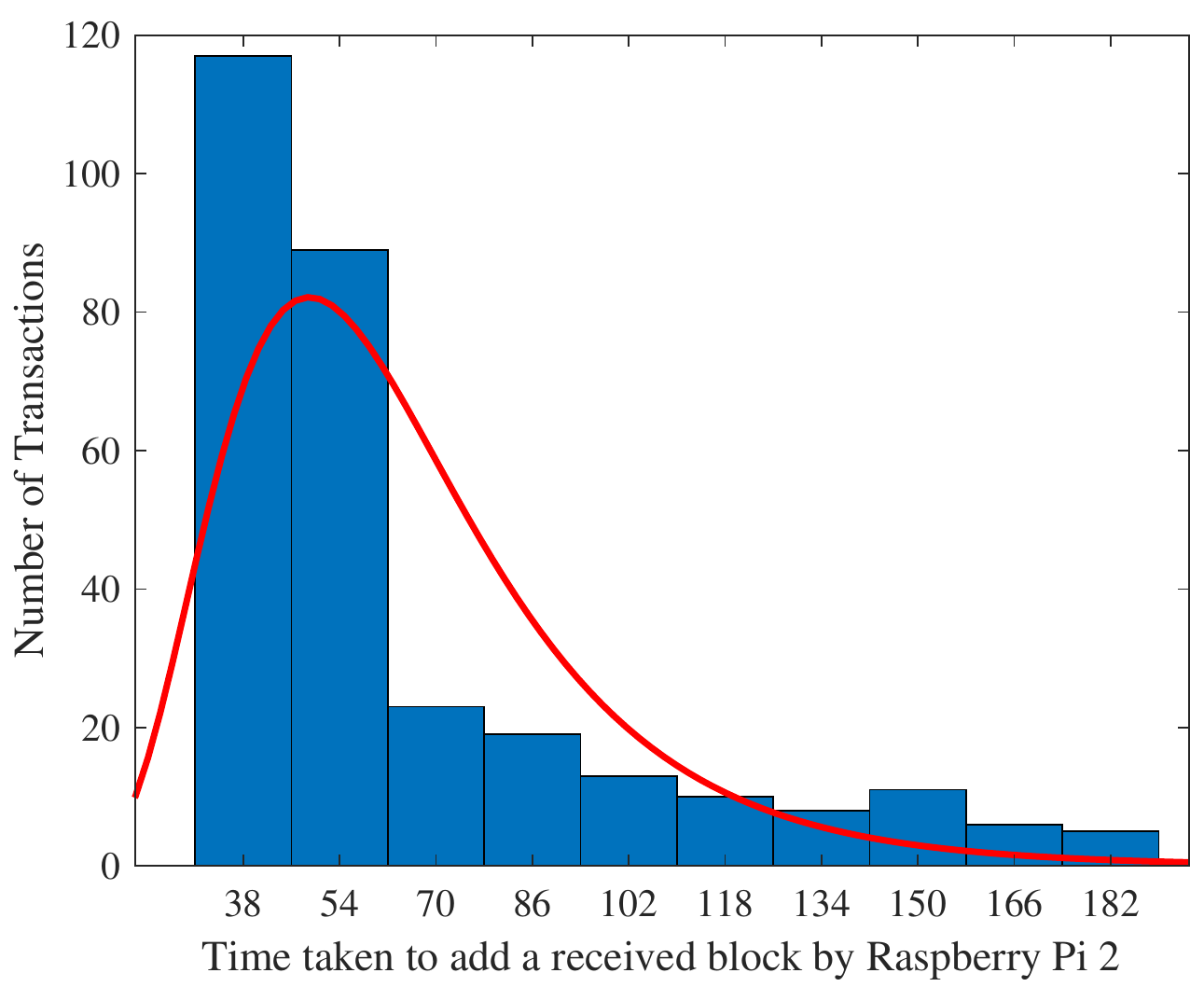}}\qquad
	\subfloat[ClearPi (Raspbery Pi 1 Model B+)]{\includegraphics[width=0.45\textwidth]{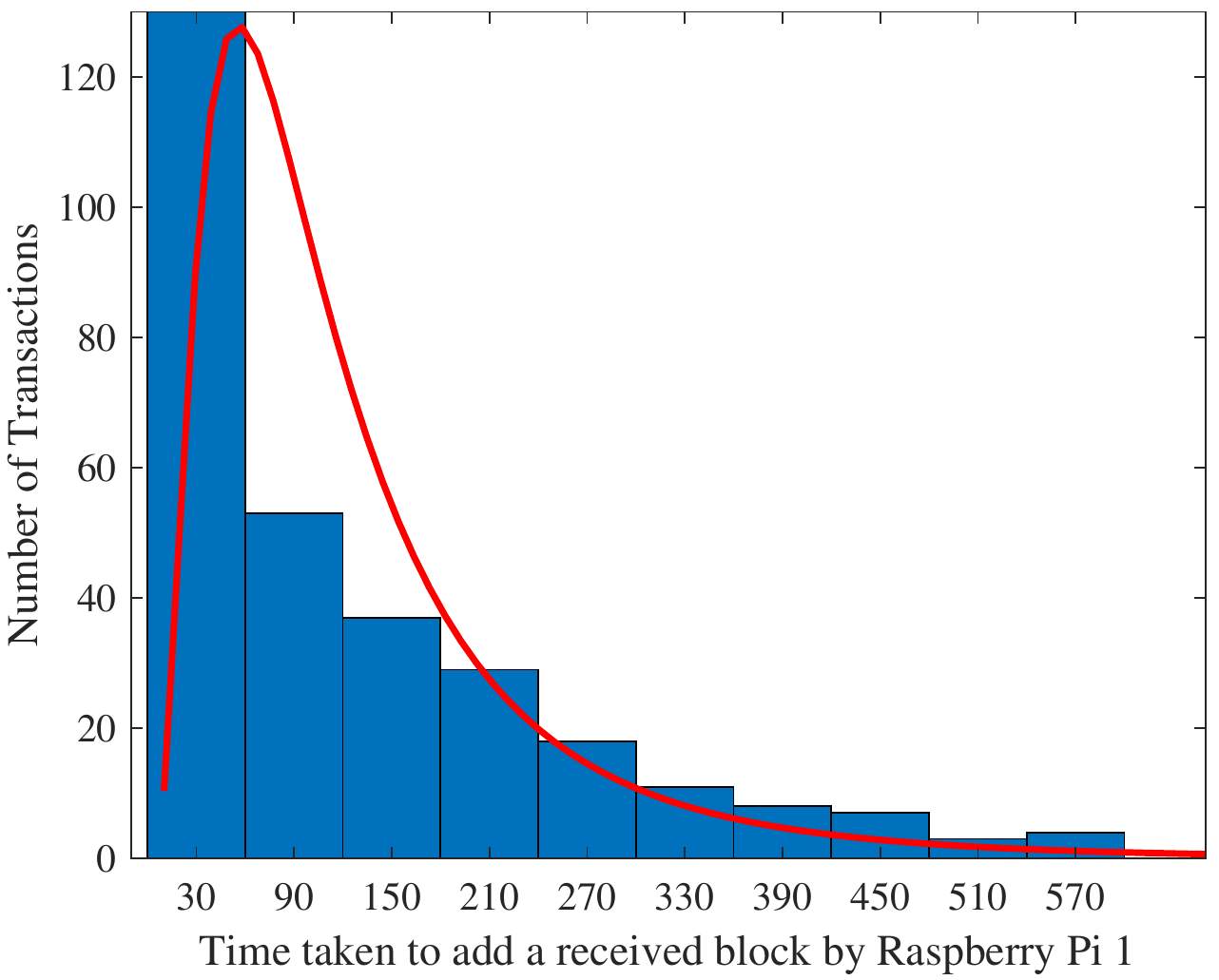}}
	\caption{Frequency plot of time required for block addition.}
	\label{FIG:Add_Histo}
\end{figure*}

\begin{figure*}[t]
	\centering
	\subfloat[BlackPi]{\includegraphics[width=0.45\textwidth]{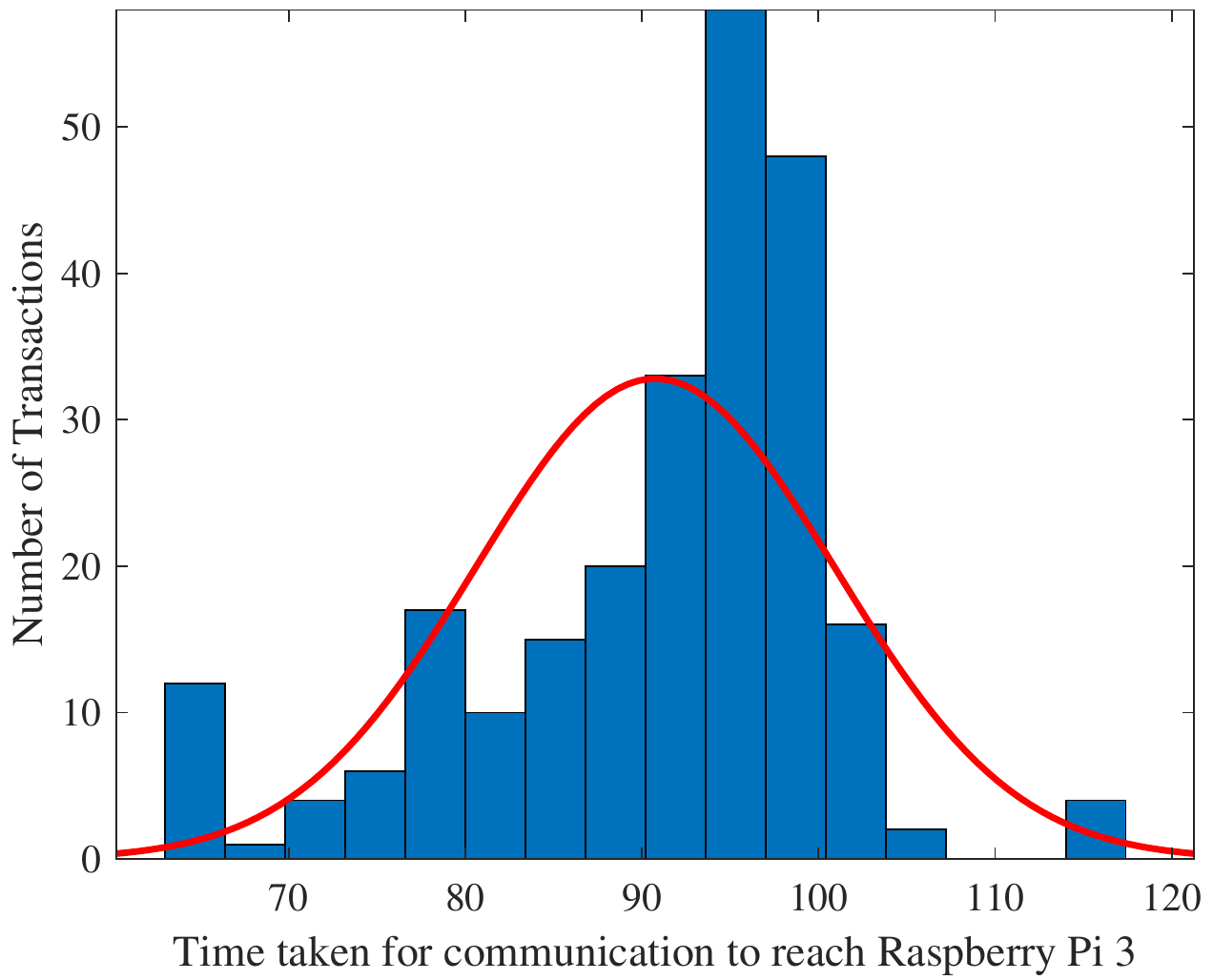}}\qquad
    \subfloat[ClearPi (Raspbery Pi 2 Model B+)]{\includegraphics[width=0.45\textwidth]{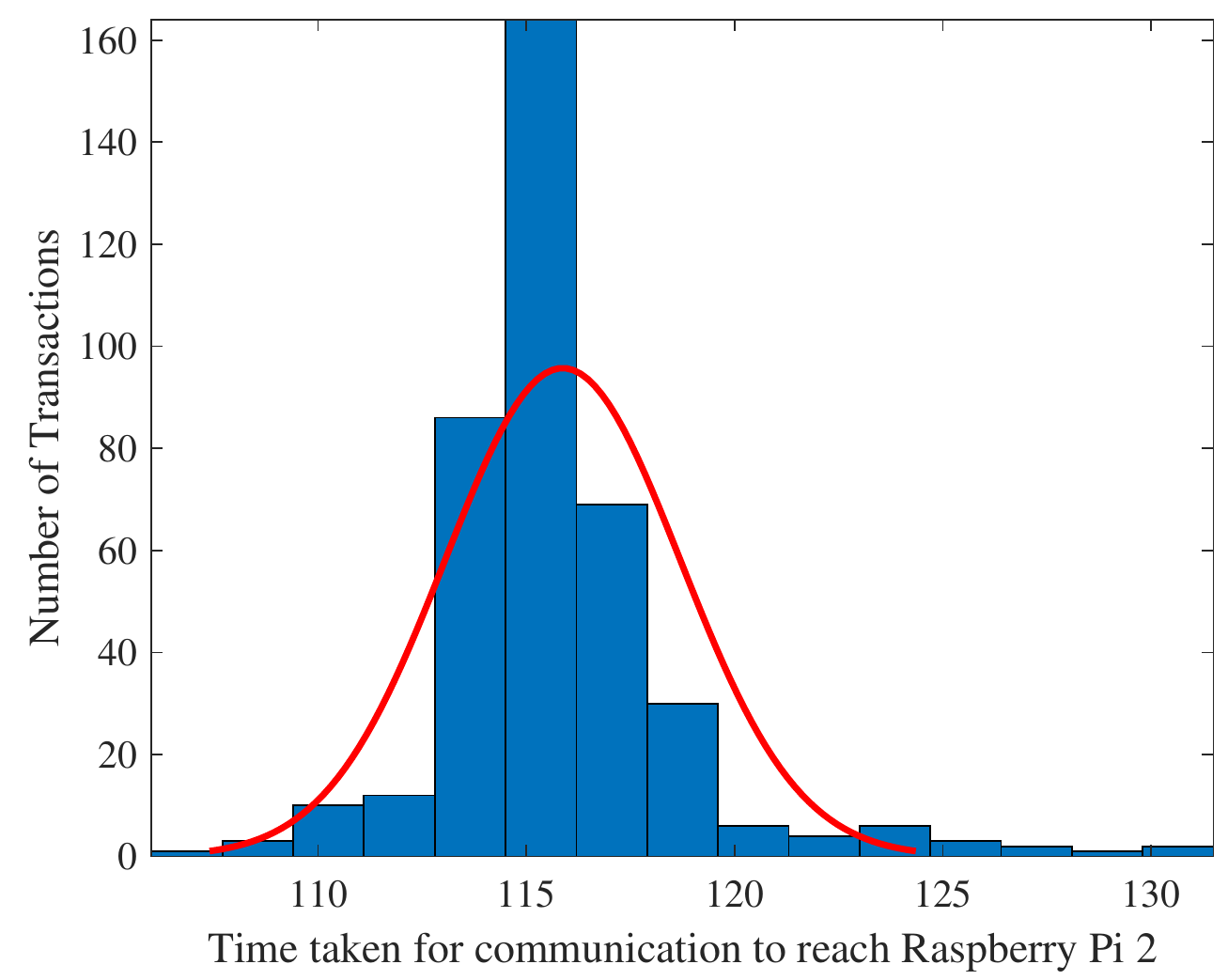}}\qquad
    \subfloat[ClearPi (Raspbery Pi 1 Model B+)]{\includegraphics[width=0.45\textwidth]{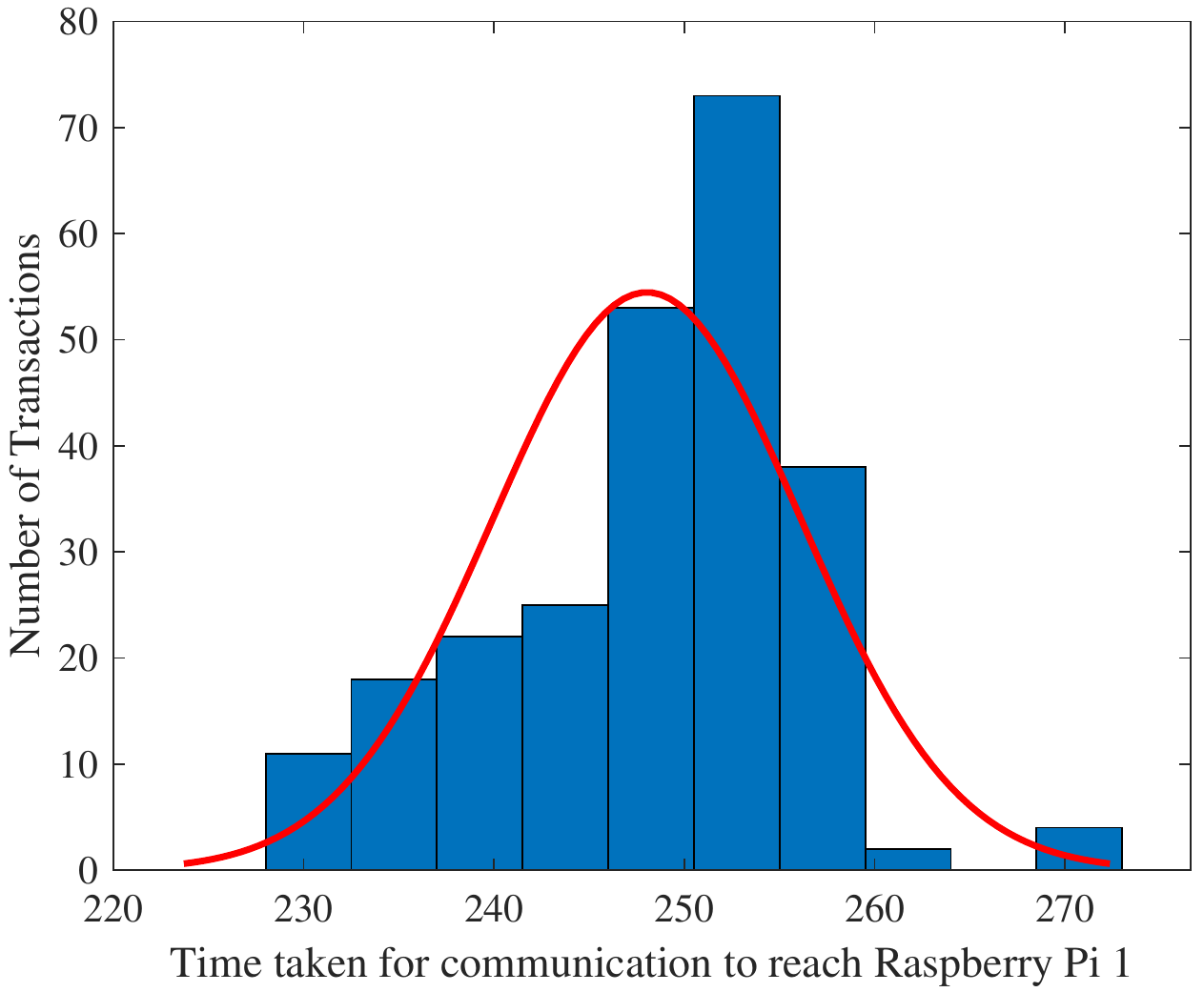}}
	\caption{Frequency plot of time required for communication.}
	\label{FIG:Comm_Histo}
\end{figure*}

Fig. \ref{fig:Power_Consumption} shows the power consumption results. The minimum power consumption is when the Raspberry Pis are waiting for the data to arrive. In the case of a client node, they are collecting the data and transmitting at a 10 sec interval. The minimum power consumption is when the nodes are neither collecting the data nor they are adding the blocks to the blockchain. The maximum power consumption is when the CPU usage is at its maximum which indicates that the nodes are collecting the data and adding a received block to the blockchain. \textcolor{black}{Fig. \ref{fig:Power_Consumption} results are abstracted from the testbed as shown in Fig. \ref{FIG:Experimental_Setup_Power_Consumption} and the energy consumption is listed in Table \ref{TBL:POWER_CONSUMPTION}.} In the case of a miner or validator node, the minimum power consumption is when it is idle, i.e. when no communication reaches the node and it is not adding any new blocks to the blockchain. The maximum power consumption is when the block is being validated, added to the blockchain and broadcast to the rest of the network. The times required to add blocks are tabulated in Table \ref{TBL:Adding_Blocks_To_Blockcahin} while the times required for communication are tabulated in Table \ref{TBL:Communication_Between_Devices}. The frequency plots of the measured results for the block addition and communication are shown in Fig. \ref{FIG:Add_Histo} and Fig. \ref{FIG:Comm_Histo}, respectively. A comparative analysis of the proposed PoAh algorithm to other consensus algorithms is given in Table \ref{TBL:Comparison}, \textcolor{black}{in which, we have presented the mining process, possible attacks and power consumption in comparison to PoAh. It is clear that the PoW mining process took around 538 KW energy and is impossible to deploy in recourse constraints IoT devices. Further we tested PoS, which is cutting the energy consumed by a hundredfold (i.e. around 99\%). As a result, the PoS mining process is taking around 5.5 KW and IoT applications sill demands novel blockchain consensus for resource constraint devices. Compared to the above most promising blockchain consensus, the proposed PoAh is introducing authentication mechanisms and taking a maximum of 3.5 Watts energy for block authentication. PoAh is much faster compared to PoW and PoS, and suitable for IoT devices, which is experimented and validated in the in-lab hardware testbed.}

\begin{table}[htbp]
\caption{\textcolor{black}{POWER CONSUMPTION OF DIFFERENT RASPBERRY Pis.}}
	\label{TBL:POWER_CONSUMPTION}
	\centering
	\begin{tabular}{|p{5.2cm}p{2.6cm}p{2.6cm}p{2.6cm}|} 
		\hline
\textbf{Power Types}		 & \textbf{Raspberry Pi 1} & \textbf{Raspberry Pi-2} & \textbf{Raspberry Pi-3} \\
		\hline 
\hline
		Max Power consumption in Watts & 1.8 & 2.5 & 3.6 \\
		\hline
		Min Power consumption in Watts & 1.5 & 2 & 3.1 \\
		\hline
	\end{tabular}
\end{table}

\begin{table}[htbp]
\caption{TIME TAKEN BY PoAh FOR ADDING THE BLOCKS RECEIVED.}
	\label{TBL:Adding_Blocks_To_Blockcahin}
	\centering
	\begin{tabular}{|p{5.7cm}p{2.5cm}p{3.7cm}|} 
		\hline
\textbf{Single Board Computer} & \textbf{Mean (Milliseconds)} & \textbf{Standard Deviation (Milliseconds)} \\
		\hline 
\hline
		BlackPi & 843 & 26 \\
		\hline
		ClearPi (Raspberry Pi 2 Model B+) & 85 & 36\\
		\hline
		ClearPi (Raspberry Pi 1 Model B+) & 162.4 & 98.6\\
		\hline
	\end{tabular}
\end{table}

\begin{table}[htbp]
\caption{TIME TAKEN FOR COMMUNICATION.}
	\label{TBL:Communication_Between_Devices}
	\centering
	\begin{tabular}{|p{5.7cm}p{2.5cm}p{2.4cm}|}
		\hline
\textbf{Single Board Computer} & \textbf{Mean (Milliseconds)} & \textbf{Standard Deviation (Milliseconds)} \\ [0.5ex] 
		\hline
		\hline
		BlackPi & 89 & 20 \\
		\hline
		ClearPi (Raspberry Pi 2 Model B+) & 116.35 & 12.3 \\
		\hline
		ClearPi (Raspberry Pi 1 Model B+) & 246 & 33\\
		\hline
	\end{tabular}
\end{table}

\begin{table*}[t]
	\caption{\textcolor{black}{COMPARATIVE PERSPECTIVE OF PoAh WITH OTHER RELATED WORKS.}}
	\label{TBL:Comparison}
	\centering
	\begin{tabular}{|p{3.0cm}cp{2.5cm}p{2.2cm}p{2.0cm}p{2.5cm}|}
		\hline
		\textbf{Consensus Algorithm} & \textbf{Year} & \textbf{Blockchain Type} & \textbf{Mining}  &  \textbf{Prone To Attacks} & \textbf{Power Consumption} \\ [0.5ex] 
		\hline
		\hline
		Proof-of-Work \cite{Nakamoto_bitcoin}&2008&Permission-Less&Based on Computation Power&Bribe attack, Sybil attack, 51\% attack&538 KWh Electricity consumption \\
		\hline
		Proof-of-Stake \cite{PoS}&2012&Permission-Less&Validation& DoS, Sybil attack & 5.5 KWh Electricity consumption\\
		\hline
		Ripple \cite{ripple}&2014 & Permissioned & Vote Based Mining & DoS attack, Sybil attacks& -\\
		\hline
		Proof-of-Vote \cite{Li_PoV}&2017 & Consortium&Vote Based Mining& - &-\\
		\hline
		Proof-of-Trust \cite{Zou_PoT} & 2018 & Permission Based &Probability and	Vote Based Mining &DDoS attack& - \\
		\hline
Proof  of PUF-Enabled Authentication (PoP) \cite{Mohanty_arXiv_2019-Sep17-1909-06496_PUFchain,Mohanty_IEEE-MCE_2020-Mar_PUFchain} & 2019 & Permission Based & Authentication & No Known Attacks&  5 W (Ultra Low Power Consumption) \\		
		\hline
		Proof-of-Authentication (PoAh) \textbf{(The Current Paper)} & 2019 & Permission Based & Authentication & No Known Attacks& 3.5 W (Ultra Low Power Consumption) \\
		\hline
	\end{tabular}
\end{table*}

\section{Conclusions}
\label{Sec:Conclusions}

This paper provides a novel consensus algorithm named Proof-of-Authentication (PoAh) for sustainable and lightweight blockchain for resource-constrained distributed systems, such as IoT and edge computing. Furthermore, the proposed algorithm bypasses centralized dependencies by building a lightweight decentralized security solution. The proposed algorithm is validated in terms of security and sustainability in three steps: (i) theoretical validation, (ii) simulation results, and (iii) real-time test-bed deployment. The proposed PoAh, while running in
limited computer resources and using minimal energy (e.g. single-board computing devices like the Raspberry Pi) has a latency in the order of few secs. Thus PoAh is scalable for largescale IoT deployed in the smart cities. In future work, the PoAh algorithm will be evaluated in big data scenarios for large scale networks. We are also planning a comparative study of various benchmark security solutions in large scale networks, \textcolor{black}{as well as evaluation of various threat attack scenarios.}

\section{Acknowledgment}

A preliminary version of this study has been presented at following \cite{Puthal_Potentials_2019-Jan,Puthal_ICCE_2019_PoAh}.

The authors would like to sincerely thank Dr. Gautam Das for his feedback on conference version of this work.

\bibliographystyle{IEEEtran}



\pagebreak
\section*{Authors' Biographies}

\begin{minipage}[htbp]{\columnwidth}
	\vspace{1.0cm}	
	\begin{wrapfigure}{l}{1.4in}
		\vspace{-0.5cm}
		\includegraphics[width=1.4in,keepaspectratio]{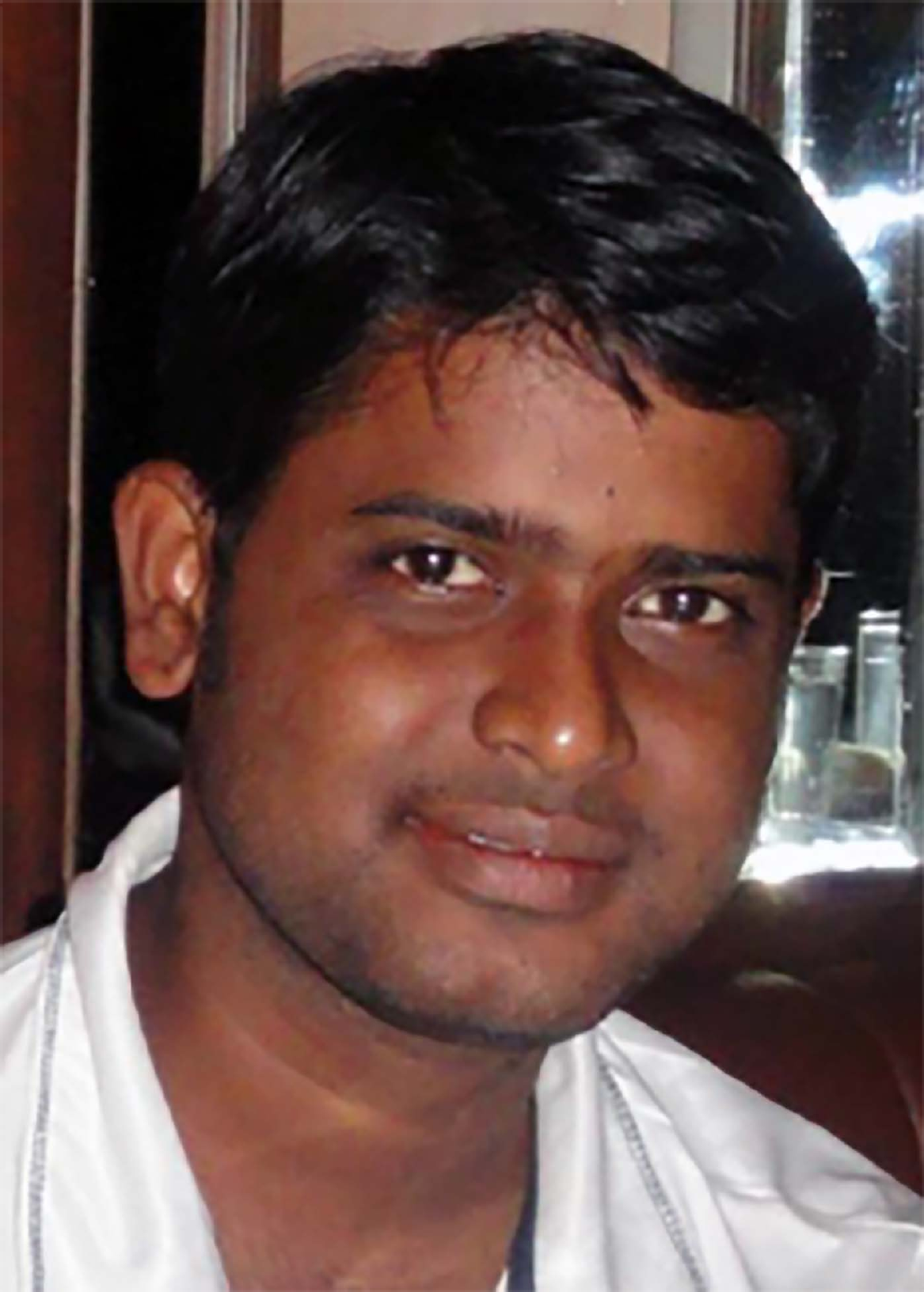}
		\vspace{-0.9cm}
	\end{wrapfigure}
	\noindent
	\textbf{Deepak Puthal} (M'16) received the Ph.D. degree in computer science from the University of Technology Sydney (UTS), Australia. He is currently a Lecturer at School of Computing, Newcastle University, Newcastle upon Tyne, UK. He is an author/co-author of more than 100 peer-reviewed publications in international conferences and journals, including ACM and IEEE transactions. His research interests include cyber security, Internet of Things, distributed computing, and edge/fog computing. He has been a Program Chair and a Program Committee member in several IEEE and ACM sponsored conferences. He was a recipient of the 2017 IEEE Distinguished Doctoral Dissertation Award from the IEEE Computer Society and STC on Smart Computing. He served as a Co-Guest Editor of several reputed journals, including Concurrency and Computation: Practice and Experience, Wireless Communications and Mobile Computing, and Information Systems Frontier. He is an Associate Editor of the IEEE Transactions on Big Data, and IEEE Consumer Electronics Magazine.
\end{minipage}

\vspace{1.0cm}

\begin{minipage}[htbp]{\columnwidth}
	\begin{wrapfigure}{l}{1.40in}
		\vspace{-0.5cm}
		\includegraphics[width=1.40in,keepaspectratio]{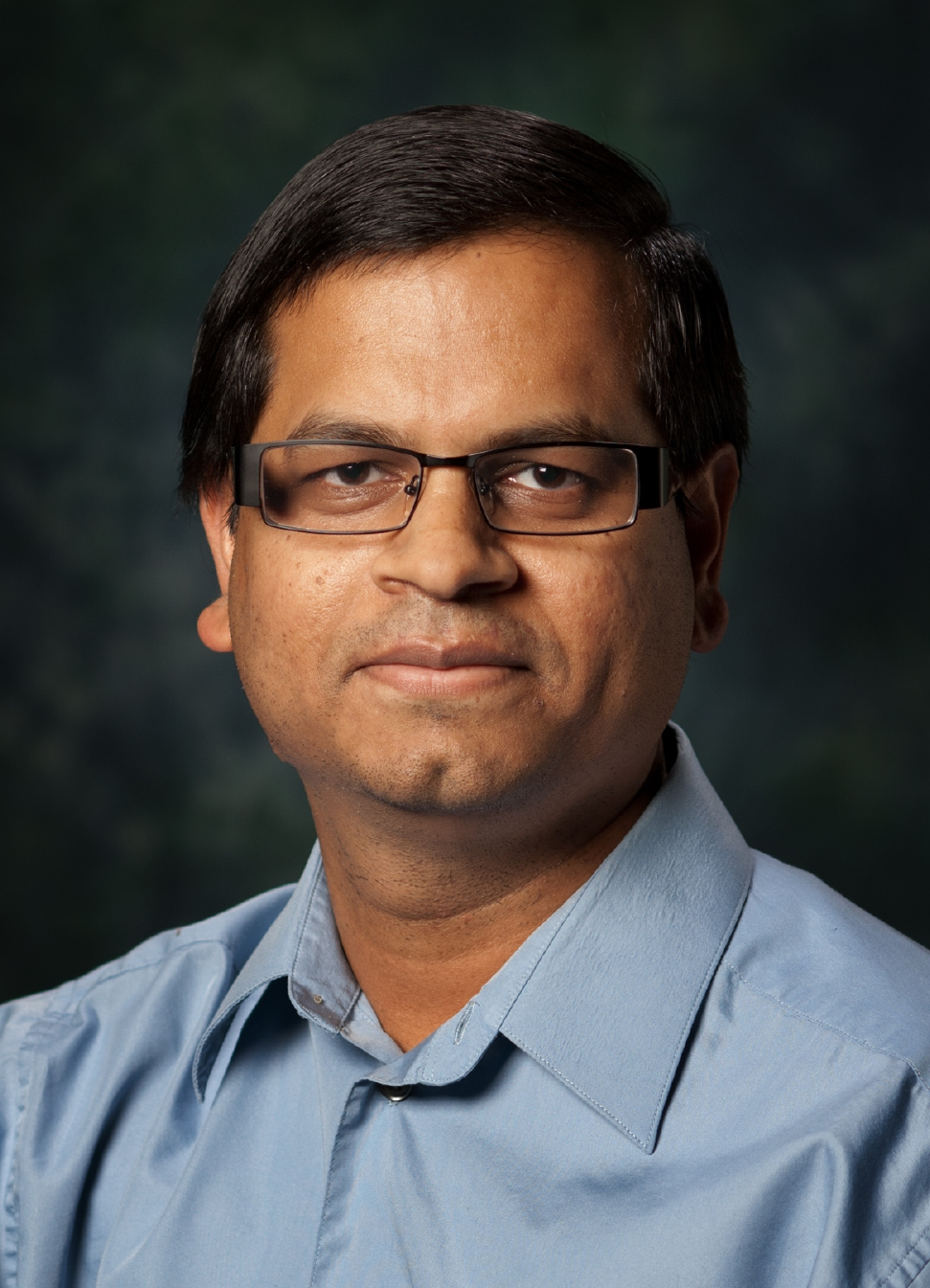}
		\vspace{-0.9cm}
	\end{wrapfigure}
	\noindent
\textbf{Saraju P. Mohanty} (SM'08) received the bachelor's degree (Honors) in electrical engineering from the Orissa University of Agriculture and Technology, Bhubaneswar, in 1995, the master's degree in Systems Science and Automation from the Indian Institute of Science, Bengaluru, in 1999, and the Ph.D. degree in Computer Science and Engineering from the University of South Florida, Tampa, in 2003.
He is a Professor with the University of North Texas. His research is in ``Smart Electronic Systems'' which has been funded by National Science Foundations (NSF), Semiconductor Research Corporation (SRC), U.S. Air Force, IUSSTF, and Mission Innovation Global Alliance. He has authored 300 research articles, 4 books, and invented 4 U.S. patents. His has Google Scholar citations with an h-index of 34 and i10-index of 124 with 5000+ citations. He was a recipient of 11 best paper awards, IEEE Consumer Electronics Society Outstanding Service Award in 2020 for leadership contributions, the IEEE-CS-TCVLSI Distinguished Leadership Award in 2018 for services to the IEEE and to the VLSI research community, and the 2016 PROSE Award for Best Textbook in Physical Sciences and Mathematics category from the Association of American Publishers for his Mixed-Signal System Design book published by McGraw-Hill. 
He has delivered 9 keynotes and served on 5 panels at various International Conferences. He has been serving on the editorial board of several peer-reviewed international journals, including IEEE Transactions on Consumer Electronics (TCE), and IEEE Transactions on Big Data (TBD). 
He is currently the Editor-in-Chief (EiC) of the IEEE Consumer Electronics Magazine (MCE). 
He has been serving on the Board of Governors (BoG) of the IEEE Consumer Electronics Society, and has served as the Chair of Technical Committee on Very Large Scale Integration (TCVLSI), IEEE Computer Society (IEEE-CS) during 2014-2018. He is the founding steering committee chair for the IEEE International Symposium on Smart Electronic Systems (iSES), steering committee vice-chair of the IEEE-CS Symposium on VLSI (ISVLSI), and steering committee vice-chair of the OITS International Conference on Information Technology (ICIT). He has mentored 2 post-doctoral researchers, and supervised 11 Ph.D. dissertations, 26 M.S. theses, and 10 undergraduate projects. 
\end{minipage}

\vspace{1.0cm}

\begin{minipage}[htbp]{\columnwidth}
	\begin{wrapfigure}{l}{1.40in}
		\vspace{-0.5cm}
		\includegraphics[width=1.40in,keepaspectratio]{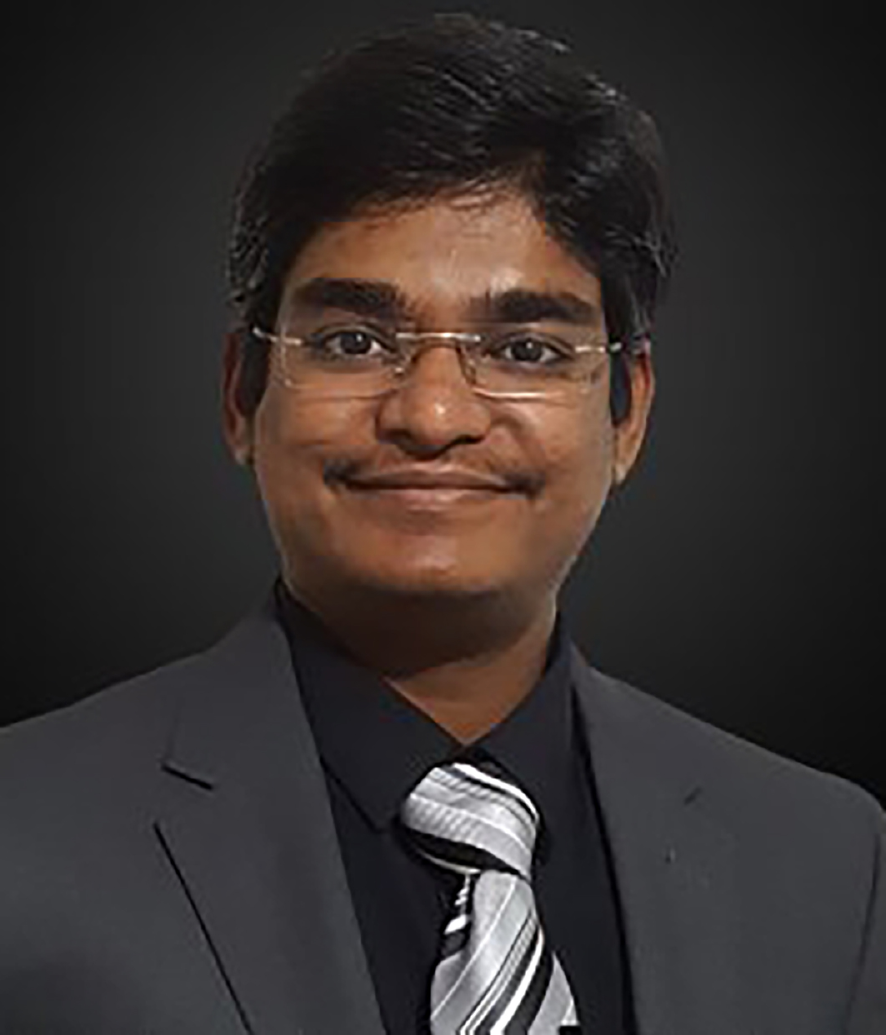}
		\vspace{-0.9cm}
	\end{wrapfigure}
	\noindent 
	\textbf{Venkata P. Yanambaka} (M'19) received the Bachelor of Technology degree in electronics and communications from the JNTU, India, in 2014.  He obtained his Ph.D. at the System Electronic Systems Laboratory (SESL) at the Department of Computer Science and Engineering, University of North Texas. He is currently an Assistant Professor in the School of Engineering and Technology, Central Michigan University. His research interests are in Security in Internet of Things (IoT), Energy-Efficient Circuits and Systems, and Application-Specific Systems Design. He has authored of a 12 research articles which include multiple journals/transactions articles. 
	He has a regular reviewer of various peer-reviewed journals and conferences.
\end{minipage}

\vspace{1.0cm}

\begin{minipage}[htbp]{\columnwidth}
	\vspace{1.0cm}	
	\begin{wrapfigure}{l}{1.4in}
		\vspace{-0.5cm}
		\includegraphics[width=1.4in,keepaspectratio]{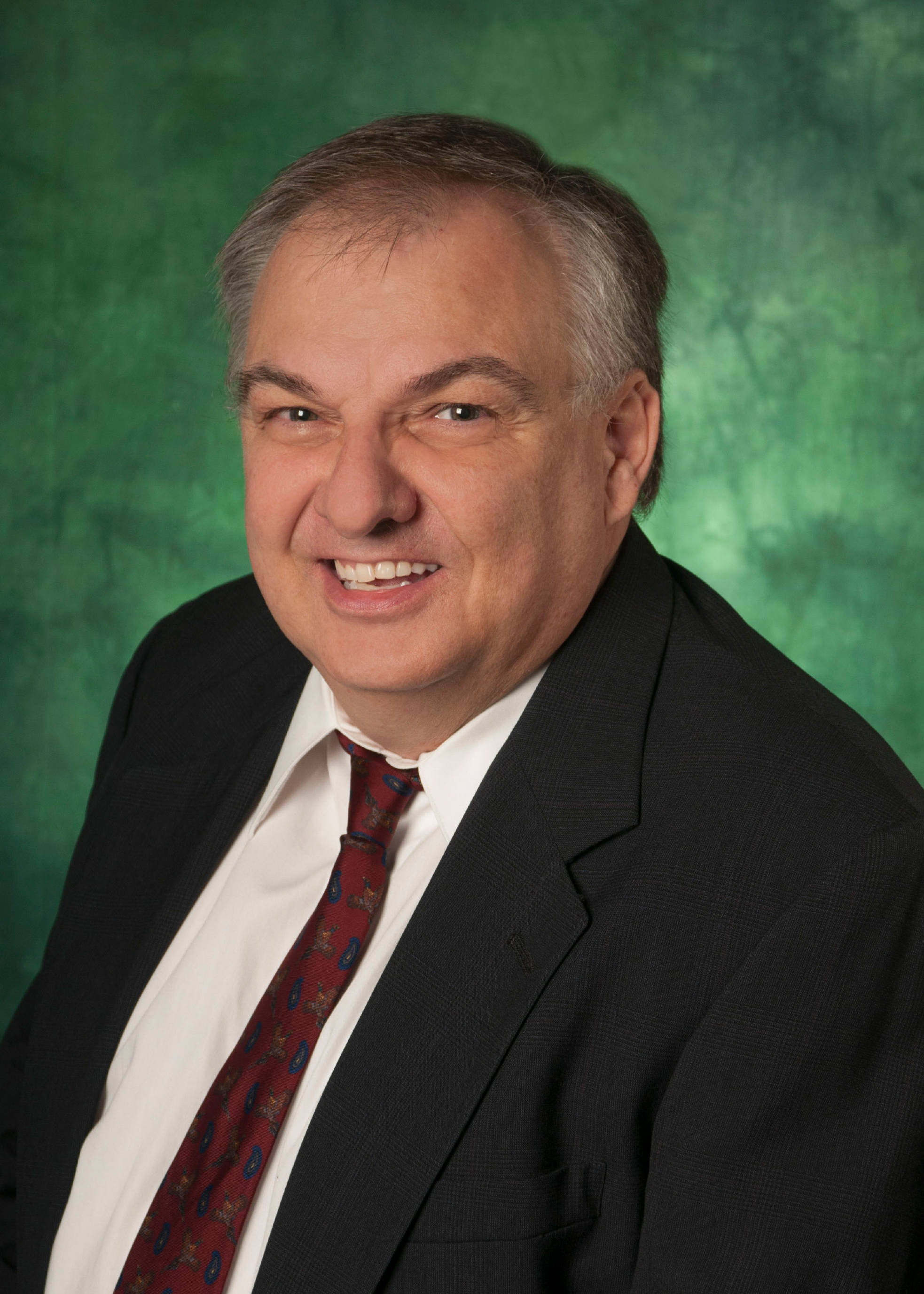}
		\vspace{-0.9cm}
	\end{wrapfigure}
	\noindent
	\textbf{Elias Kougianos} (SM'07) received a BSEE from the University of Patras, Greece in 1985 and an MSEE in 1987, an MS in Physics in 1988 and a Ph.D. in EE in 1997, all from Louisiana State University. 
	From 1988 through 1997 he was with Texas Instruments, Inc., in Houston and Dallas, TX.
	Initially he concentrated on process integration of flash memories and later as a researcher in the areas of Technology CAD and VLSI CAD development. 
	In 1997 he joined Avant! Corp. (now Synopsys) in Phoenix, AZ as a Senior Applications engineer and in 2001 he joined Cadence Design Systems, Inc., in Dallas, TX as a Senior Architect in Analog/Mixed-Signal Custom IC design. He has been at UNT since 2004. He is a Professor in the Department of Electrical Engineering, at the University of North Texas (UNT), Denton, TX. His research interests are in the area of Analog/Mixed-Signal/RF IC design and simulation and in the development of VLSI architectures for multimedia applications. 
	He is an author of over 120 peer-reviewed journal and conference publications.
\end{minipage}


\end{document}